\documentclass[11pt]{article}

\usepackage{multirow}
\usepackage{amssymb,amsmath,amsthm}
\usepackage{color}
\usepackage{graphicx}
  \graphicspath{{./Figures/}}
  \DeclareGraphicsExtensions{.jpeg,.png,.jpg,.eps,.pdf}
\usepackage{tikz}
\usepackage{epstopdf}
\usepackage{hyperref}

\newcommand{\bc}{\begin{center}}
\newcommand{\ec}{\end{center}}
\newcommand{\be}{\begin{equation}}
\newcommand{\ee}{\end{equation}}
\newcommand{\bd}{\begin{displaymath}}
\newcommand{\ed}{\end{displaymath}}
\newcommand{\ba}{\begin{array}}
\newcommand{\ea}{\end{array}}
\newcommand{\ben}{\begin{enumerate}}
\newcommand{\een}{\end{enumerate}}
\newcommand{\bit}{\begin{itemize}}
\newcommand{\eit}{\end{itemize}}
\newcommand{\beq}{\begin{eqnarray}}
\newcommand{\eeq}{\end{eqnarray}}
\newcommand{\btab}{\begin{tabular}}
\newcommand{\etab}{\end{tabular}}
\newcommand{\bfig}{\begin{figure}}
\newcommand{\efig}{\end{figure}}
\newcommand{\btp}{\begin{tikzpicture}}
\newcommand{\etp}{\end{tikzpicture}}

\definecolor{verm}{rgb}{0.6,0.2,0.2}
\definecolor{purp}{rgb}{0.3,0.1,0.6}
\definecolor{purple}{rgb}{0.4,0.0,0.6}
\definecolor{bggreen}{rgb}{0.1,0.3,0.1}
\definecolor{dgreen}{rgb}{0.1,0.6,0.1}
\definecolor{black}{rgb}{0.0,0.0,0.0}
\definecolor{crim}{rgb}{0.3,0.1,0.1}
\definecolor{dred}{rgb}{0.5,0.1,0.1}

{\bf}{\it}
{\bf}{\it}
{\bf}{\rm}
\newtheorem{lemma}{Lemma}{\bf}{\it}
\newtheorem{theorem}{Theorem}{\bf}{\it}
{\bf}{\it}
{\bf}{\it}
{\bf}{\rm}

\newenvironment{oldTheorem}[1]{\begin{trivlist}%
\item[\hskip\labelsep{\bf Theorem~\ref{#1}~}]\it}{\rm\end{trivlist}}
\newenvironment{oldLemma}[1]{\begin{trivlist}%
\item[\hskip\labelsep{\bf Lemma~\ref{#1}~}]\it}{\rm\end{trivlist}}

\newcommand{\dx}[2]{\frac{d#1}{d#2}}

\def\e{\epsilon}
\def\g{\gamma}
\def\r{\rho}
\def\ra{\rho_{\text actual}}
\def\rgain{\rho_{\text gain}}
\def\rloss{\rho_{\text loss}}
\def\u{{\mathbf u}}
\def\v{{\mathbf v}}
\def\w{{\mathbf w}}
\def\bt{\tilde{\beta}}
\def\rt{\tilde{\r}}

\def\N{{\mathcal N}}
\def\rN{\widehat{{\mathcal N}}}
\def\aN{\widetilde{{\mathcal N}}}
\def\R{{\mathcal R}}
\def\T{{\mathcal T}}
\def\aT{\widetilde{{\mathcal T}}}
\def\M{{\mathcal M}}
\def\C{{\mathcal C}}
\def\aC{\widetilde{{\mathcal C}}}

\def\U{{\mathcal U}}

\newcommand{\pvec}[1]{{\mathbf #1}}

\def\Md{\dot{M}}
\def\Rd{\dot{R}}
\def\Sd{\dot{S}}
\def\Td{\dot{T}}
\def\Ud{\dot{U}}

\begin{document}

\title{
SUTRA: An Approach to Modelling Pandemics with \\
Undetected Patients, and Applications to COVID-19
}

\author{Manindra Agrawal, Madhuri Kanitkar, Deepu Phillip,\\
Tanima Hajra, Arti Singh, Avaneesh Singh, \\
Prabal Pratap Singh and Mathukumalli Vidyasagar
\thanks{
MA, DP, TH, ArS, AvS, PPS are at
Indian Institute of Technology Kanpur, Kanpur, UP 208016;
MK is the Vice Chancellor of Maharashtra University of Health Sciences,
Nashik, MH 422004;
MV is with the Department of Artificial Intelligence,
Indian Institute of Technology Hyderabad, Kandi, TS 502284;
MA is the corresponding author.
Email: manindra@iitk.ac.in
}
}

\maketitle

\begin{abstract}

The Covid-19 pandemic has two key properties: (i) asymptomatic cases (both detected and undetected) that can result in new infections, and (ii) time-varying characteristics due to new variants, Non-Pharmaceutical Interventions etc. We develop a model called SUTRA (Susceptible, Undetected though infected, Tested positive, and Removed Analysis) that takes into account both of these two key properties. 

While applying the model to a region, two parameters of the model can be learnt from the number of daily new cases found in the region. Using the learnt values of the parameters the model can predict
the number of daily new cases so long as the
learnt parameters do not change substantially.
Whenever any of the two parameters changes due to the key property (ii) above, the SUTRA model can detect that the values of one or both of the parameters have changed. Further, the model has the capability to relearn the changed parameter values, and then use these to carry out the prediction of the trajectory of the pandemic for the region of concern.

The SUTRA approach can be applied at various levels of granularity, from an entire country to a district, more specifically, to any large enough region for which the data of daily new cases are available. 

We have applied the SUTRA model to thirty-two countries, covering more than half of the world's population. Our conclusions are: (i) The model is able to capture the past trajectories very well. Moreover, the parameter values, which we can estimate robustly, help quantify the impact of changes in the pandemic characteristics. (ii) Unless the pandemic characteristics change significantly, the model has good predictive capability. (iii) Natural immunity provides significantly better protection against infection than the currently available vaccines.

These properties of the model
make it useful for policy makers to plan logistics
and interventions.

\end{abstract}

\section{Introduction}

The COVID-19 pandemic caused by the SARS-CoV-2 virus has by now led to more than 
600 million reported cases and more than six million deaths worldwide,
as of October 1, 2022 \cite{worldometers}.
By way of comparison, the infuenza epidemic of 1957 led to 20,000 deaths
in the UK and 80,000 deaths in the USA, while the 1968 influenza pandemic
led to 30,000 deaths in the UK and 100,000 deaths in the USA
\cite{Honigsbaum20}.
In contrast, the COVID-19 pandemic has already led to more than
one million deaths in the USA and more than 170,000 deaths in the UK
\cite{Honigsbaum20}.
Therefore the COVID-19 pandemic is the most deadly since the Spanish Flu
pandemic which started in 1918.
In the USA, 675,000 people, or 0.64\% of the population, died in that
pandemic \cite{CDC-Spanish},
compared to 0.33\% of the population in the current pandemic.
In order to cope with a health crisis of this magnitude, governments
everywhere require accurate projections of the progress of the
pandemic, both in space and over time, and at various levels of granularity.
In addition, decision-makers also require
assessments of the relative effectiveness of different
non-pharmaceutical interventions (NPIs) such as lockdowns.

Over the past century or so, various mathematical models have been developed to predict the trajectory
of a pandemic. These can be classified in three broad categories: statistical models, state-space models, 
and empirical models~\cite{math-models}.
The most popular among these are compartment models, a type of state-space models.
These models divide population into disjoint compartments representing different stages of infection, and 
are based on the premise that the disease spreads
when an infected person comes into contact with a susceptible person.

In the initial SIR model \cite{Ker-McK27},
the population was divided into three
compartments: S (Susceptible), I (Infected), and R (Removed / Recovered).
Subsequently an intermediate compartment of E (Exposed) was introduced
between S and I \cite{Hethcote76}.
In SEIR model, interactions between S and E \textit{do not}
lead to fresh infections. For pandemics like COVID-19, that have significant number of asymptomatic patients,
instead of E, compartment A (Asymptomatic) with interactions between S and A also leading to new infections,
is more suited~\cite{Rob-Sti13}. For COVID-19 pandemic, a number of models have been introduced to capture
its trajectory. Two significant features of this pandemic are the presence of a large number of undetected cases and
time-varying parameter values due to the emergence of various mutants, lockdowns etc. For any model to  capture the trajectory well, it must
take into account these two features.
Some of the proposed models have large number of compartments in order to make the model biologically more realistic
(for example~\cite{SIDARTHE}). These models have a large number of parameters, and estimating their values reliably is not
possible from reported data due to the well-known phenomenon of the bias-variance tradeoff in statistics. This is even more true when the parameter values change with time. Besides the ones mentioned above, there are other models work with four compartments (and consequently a small number of parameters). In these models, parameter estimation is easier (for example~\cite{SAIR-AOC,SUTR-posterior,SIRD-time-varying}).
However, even these models have to make certain assumptions about how the parameter values change. For example,
\cite{SAIR-AOC} assumes the parameter values are derived from other considerations and do not change with time,
while both \cite{SUTR-posterior} and \cite{SIRD-time-varying} assume that parameters change following a specific equation.

\section{Our Contributions}

Against this background, in this paper we propose a four compartment 
model called SUTRA (Susceptible, Undetected,
Tested positive, and Removed, Approach).%
\footnote{In Sanskrit, the word Sutra also means an aphorism.
Sutras are a genre of ancient and medieval Hindu
texts, and depict a code strung together by a genre.} The components are same as in~\cite{SUTR-posterior}, however,
there are differences in the dynamics (with better epidemiological justification, see~\ref{sec-model}).
This has following consequences.

\subsection{Fundamental Equation}

The model admits a time invariant relationship (called {\em fundamental equation}) between detected new cases ($\N_T$), detected active cases ($\T$), and total detected cases ($\C_T$): 
\[ \T(t) = \frac{1}{\bt} \N_T(t+1) + \frac{1}{\rt P_0} \C_T(t) \T(t) \]
where $\bt$ and $\rt$ are parameters of the model and $P_0$ the population of region under study (see section~\ref{sec-model}). 

\subsection{Estimation of $\bt$ and $\rt$}

We can efficiently estimate values of the two parameters $\bt$ and $\rt$ for the entire duration of the pandemic using the
fundamental equation and simple linear regression (see section~\ref{sec-detected-parameters}). These parameters have the following interpretation:

\begin{itemize}
\item
Parameter $\bt \approx \beta$, a standard parameter denoting contact rate (also called transmission rate). 
\item
We may interpret $\rt \approx \e$ where $\e$ is {\em detection ratio}, the ratio of detected new cases to total number of new cases (as done in an earlier version of our model~\cite{MV-IJMR20}), however, it is not satisfactory since in every region, value of $\rt$ is observed to increase by a very large factor ($1000$ or more) in the initial couple of months of pandemic before becoming much less volatile (see section~\ref{sec-past}). A better interpretation is 
$\rt \approx \epsilon\rho$ where $\rho P_0$ is effective population under the pandemic influence (see section~\ref{sec-model}). The large initial increase then makes sense since the effective population under pandemic at the beginning is a very small fraction of $P_0$ and increases very rapidly.
\end{itemize}

\subsection{Phases of Pandemic}

The value of parameter $\bt$ reduces when restrictive measures like lockdowns are imposed, and increases when these measures are lifted or a more infectious mutant arrives.
Change in the value of parameter $\rt$ happens for many reasons. For example, when pandemic spreads to newer regions, or is completely
eliminated from a region, or a part of susceptible population gets vaccine-induced immunity, or a part of population
with immunity (acquired through vaccination or prior infection) loses it (see section~\ref{sec-phases}).

When the value of $\bt$ or $\rt$ changes significantly, the trajectory of the pandemic changes.
The model captures it as a {\em phase change} and recomputes the new values. As explained in 
section~\ref{sec-phases},
the model can detect when the values of $\bt$ or $\rt$ are changing, and when do they stabilize.

\subsection{Future Projection}

With the knowledge of $\rt$ and $\bt$, model can efficiently compute trajectory of the pandemic for the entire duration. The computed trajectory is a good estimate for future also as long as parameters do not change significantly (see section~\ref{sec-future}).

\subsection{Estimation of $\r$ and $\e$}

To understand the impact of pandemic better, it is desirable to estimate values of
$\r$ and $\e$ separately instead of their product. We show (Theorem~\ref{t-finite-eps}) that given values of $\bt$ and $\rt$ along with number of total and active infections on starting date of the simulation, there are only finitely many possible values for $\r$ and $\e$. Further, there is a {\em unique canonical value} for both
at each time instant.

Since infections count at the start
of simulation is not known, one requires a more realistic condition to be able to compute $\rho$ and $\epsilon$ values. Towards this, we show that the condition can be replaced by knowledge of value of $\epsilon$ or $\rho$ at any one time instant
(section~\ref{sec-all-parameters}).
Former requirement is met with a good serosurvey of the region at any point in time. Latter requirement is achievable if we can identify
the time when pandemic has spread all over the region making $\rho$ close to $1$. One also requires that there is
little vaccine-induced immunity at the time, since vaccine immunity reduces $\r$ (see Lemma~\ref{l-gain-loss}).

For COVID-19, as the Omicron mutant arrived nearly eighteen months after pandemic started and
is supposed to have bypassed vaccine-immunity nearly completely (as we also show in section~\ref{sec-immunity}), we can assume $\rho \approx 1$ sometime after Omicron reached a region that did not implement strict control measures at the time.
For such regions (that cover almost the entire world barring exceptions like China), we can estimate values of $\r$ and $\e$.

\subsection{Analysis of the Past} 

The computed parameter values provide a quantification of impact of various events during the course
of the pandemic. This includes impact of lockdowns and other restriction measures and arrival of new mutants. Section~\ref{sec-past} does this analysis for four countries.

\subsection{Analysis of Immunity Loss}

The Omicron mutant caused widespread loss of immunity. Applying the model on thirty-two
countries covering all continents and more than half the world's population, we deduce that loss of vaccine-immunity was
significantly more than natural immunity conferred by prior exposure to any variant (see section~\ref{sec-immunity}).

This, coupled with the fact that vaccines continue to protect against severe infection, strongly suggests that the best strategy
to manage the pandemic is to allow it to spread after vaccinating the population.

\section{Model Formulation}
\label{sec-model}

Perhaps the earliest paper to propose a pandemic model incorporating
asymptomatic patients is \cite{Rob-Sti13}.
In this paper, the population is divided into four compartments:
$S$, $A$ (for Asymptomatic), $I$ (for Infected) and $R$.
Interactions between members of $S$ and $A$, as well as between
members of $S$ and $I$, can lead to fresh infections.
In that paper, it is assumed that almost all persons in $A$ escape
detection, while almost all persons in $I$ are detected by the health
authorities.
While the SAIR model of \cite{Rob-Sti13} is a good starting point
for modeling diseases with asymptomatic patients, and has been used in a few models for COVID-19 (\cite{SAIR-AOC} for
example), it is not a good fit for COVID-19 for the following reasons:
(i) due to contact tracing, some fraction of $A$ does get detected and is often of similar order as
detected symptomatic ones, (ii) many symptomatic cases are not detected. Therefore, the size of $I$ cannot be estimated well.

In the present paper we propose a different grouping,
namely:
$S$ = Susceptible Population,
$U$ = Undetected cases in the population, 
$T$ = Tested Positive, either asymptomatic or symptomatic, and 
$R$ = Removed, either through recovery or death.
This leads to the SUTRA model, where
the last A in SUTRA stands for ``approach.'' 
Same division is used for models in~\cite{SUTR-posterior,SIRD-time-varying}.
As is standard, we use symbols $S$, $U$, $T$, $R$ to also represent (time varying) fractional size of the four compartments.

The category $R$ of removed can be further subdivided into $R_U$ denoting
those who are removed from $U$, and $R_T$
 denoting those who are  removed from $T$. 
As in the conventional SAIR model \cite{Rob-Sti13},
interactions between members of $S$ on the and members of $U$ or $T$,
can lead to the person in S getting infected with a certain likelihood.

\bfig[h]
\bc
\btp[line width=2pt]


\draw (0,0) rectangle (1,1) node [pos=0.5] {$S$} ;
\draw (4,0) rectangle (5,1) node [pos=0.5] {$U$} ;
\draw (8,0) rectangle (9,1) node [pos=0.5] {$T$} ;
\draw (11,0) rectangle (12,1) node [pos=0.5] {$R$} ;


\draw [->] (1,0.5) -- node [above] {$\beta SU$} (4,0.5) ;
\draw [->] (5,0.5) -- node [above] {$\e\beta SU$} (8,0.5) ;
\draw [->] (9,0.5) -- node [above] {$\g T$} (11,0.5) ;
\draw [->] (4.5,1) -- (4.5,2) -- node [above] {$\g U$} (11.5,2) -- (11.5,1) ;

\etp
\ec
\caption{Flowchart of the SUTRA model}
\label{fig:1}
\efig

A compartmental diagram of the SUTRA model is shown in Figure \ref{fig:1}.
Typically, to handle undetected cases, models assume that the size of $T$ is $\epsilon$ fraction of size of $U+T$
(see for example,~\cite{SUTR-posterior,SIRD-time-varying}). This is essentially equivalent to the assumption that detected new cases are $\epsilon$ fraction of new infections, as assumed in our model. Epidemiologically, all the new cases (but for rare exceptions) will remain undetected
for a few days (until the symptoms appear). Therefore, one needs to justify the choice of $\e\beta SU$ for detected new
cases. We argue as follows:
\begin{itemize}
\item
Recently infected persons have higher chances of
getting detected for two reasons. For symptomatic cases, the symptoms appears within a few days. For asymptomatic cases,
they are detected through contact tracing which mostly starts with a symptomatic case and the asymptomatic cases detected 
would all be infected {\em after} the initiating symptomatic case.
\item
Number of new cases do not change dramatically
over a few days and so number of detected cases over past few days can be taken to be proportional to
$\beta SU$, number of most recent cases.
\end{itemize}
A few additional reasonable assumptions have been made
to simplify parameter estimation.
Specifically,
\bit
\item It is assumed that the removal rate for both compartments
$T$ and $U$ is the same.
This can be justified because, due to contact tracing, a significant fraction of patients in $T$ are asymptomatic, 
and those people recover at the same rate as the asymptomatic people in $U$.
Even for the small fraction in $T$ who develop complications and pass away,
the time duration is very close to that of those who recover.
\item There is no interaction shown between the $T$ and $S$ compartments.
In most countries, those who test positive (whether symptomatic or not)
are either kept in institutional quarantine, or told to self-quarantine.
In reality, there  might still be a small amount of contact between $T$
and $S$.
However, neglecting this does not significantly change the dynamics of
the model, and greatly simplifies the parameter estimation.
\eit

With these considerations, the governing equations for the SUTRA model are:
\be\label{eq:S}
\Sd = - \beta SU , 
\ee
\be\label{eq:UT}
\Ud = \beta SU - \e \beta SU - \g U , \Td = \e \beta SU - \g T,
\ee
\be\label{eq:R}
\Rd_U = \g U , \Rd_T = \g T .
\ee
Since these quantities denote the \textit{fraction of the
population} within each compartment, we have
\bd
S + U + T + R_U + R_T = 1.
\ed
There are three parameters in above equations, namely 
$\beta$, $\g$, and $\e$. The interpretation of these parameters is as follows:
\bit
\item $\beta$ = The expected number of susceptible persons infected by an infected person in one day; it is called the {\em contact rate} or {\em transmission rate}.
\item $\g$ = {\em Removal rate}, the rate at which infected people are removed including both recoveries and deaths.
\item $\e$ = Rate at which infected patients in $U$ move over to $T$.
As shown later, it also equals the ratio $T/(U+T)$ most of the time, and is thus called the
{\em detection rate}.
\eit
Later, we introduce two more parameters $\r$ and $c$, and derive expressions for $\bt$ and $\rt$ in terms of $\beta$, $\e$,
$\r$, and $c$.

\subsection{Analyzing Model Equations} \label{sec-analysis}

Defining $M = U+T$, $R = R_U+R_T$, we get from equations~(\ref{eq:UT})
and~(\ref{eq:R}) that
\be\label{eq:MRd}
\Md + \Rd = \beta SU = \frac{1}{\e} (\Td + \Rd_T),
\ee
resulting in
\be\label{eq:MR}
M + R = \frac{1}{\e}(T + R_T) + c
\ee
for an appropriate constant of integration $c$.
Adding equations~(\ref{eq:UT}) gives
\bd
\Md = \beta SU - \gamma M = \frac{1}{\e} (\dot{T} + \gamma T) - \gamma M,
\ed
or
\be\label{eq:Md}
\dx{(M e^{\g t})}{t} = \frac{1}{\e}\dx{(T e^{\g t})}{t},
\ee
resulting in
\be\label{eq:M}
M = \frac{1}{\e} T + d e^{-\gamma t}
\ee
for some constant $d$.
Since $e^{-\gamma t}$ is a decaying exponential,
it follows that, except for an initial transient period, the relationship
$M = \frac{1}{\e} I$ holds.
This in turn implies that $U = M - T = \frac{1-\e}{\e}T$.

How long is the transient period? Observe that the constant $d$ equals $M(0) - \frac{1}{\e} T(0)$ which is close to zero
since fraction of infected cases at the start of pandemic is very small.
Therefore the transient period will not last more than a few days. As we will see later, such transient periods will recur at various stages of pandemic and all of them remain small.

Define $N_T = \Td + \Rd_T = \e\beta SU$, the fraction of population detected to be positive at time $t$, and
$C_T = T + R_T$, the fraction of population detected to be infected up to time $t$.
The above simplifications allow us to rewrite equation~(\ref{eq:UT}) as:
\be\label{eq:NT}
\begin{split}
N_T  & = \e\beta SU  = \beta(1-\e)ST \\
& = \beta(1-\e) (1 - (M+R))T \\
& = \beta(1-\e)(1 - \frac{1}{\e}(T+R_T) - c)T \\
&  = \beta(1-\e)(1-c) T - \frac{\beta(1-\e)}{\e} C_T T
\end{split}
\ee

Rearrange \eqref{eq:NT} as
\be\label{eq:NTa}
T = \frac{1}{\bt} N_T + \frac{1}{\e(1-c)}C_T T ,
\ee
where
\bd
\bt = \beta ( 1-\e )( 1-c ) .
\ed

\subsection{Discretization of the Model Relationships}

The progression of a pandemic is typically reported via two daily statistics:
The number of people who test positive, and the number of people who are
removed (including both recoveries and deaths).
The second statistics has a problem though: there is no agreement on when to classify an infected person as removed.
Some do it when RTPCR test is negative, some do it when symptoms are gone for a certain period, and some others
do it after a fixed period of time. For the purpose of modeling, this classification needs to be done at the time when
an infected person is no longer capable of infecting others. This is hard to decide, and so is almost never done.
Further, some countries do not report second statistics at all (UK for example).
In such a situation, we cannot rely on reported data, and instead compute $R_T$ by fixing $\g$ to an appropriate value
as discussed in section~\ref{sec-gamma}.

Let $\T(t)$ denote the number of active detected cases on day $t$,
$\R_T(t)$ denote the number of detected cases that are removed on or before day $t$, and $\N_T(t)$ denote the number of cases detected on day $t$.
Note that all three are integers, and $t$ is also a discrete 
counter.
In contrast, in the SUTRA model, $T$, $R_T$ and $N_T$ are 
\textit{fractions} in $[0,1]$, while $t$ is a continuum.
Therefore,
\bd
\T(t) = P\int_{t-1}^t T(s) ds , \R_T = P\int_{t-1}^t R_T(s) ds, \N_T = P\int_{t-1}^t N_T(s) ds
\ed
where $P$ is the \textit{effective population} that is potentially
affected by the pandemic.
Now we introduce the parameter measuring the spread of the pandemic.
Define number $\r$, called the {\em reach}, which equals $P/P_0$,
where $P$ is the effective population and $P_0$ is the \textit{total
population} of the group under study, e.g., the entire country, or an
individual state, or a district (this parameter is also introduced and studied in~\cite{SIRD-time-varying}).
The reach parameter $\r$ is usually nondecreasing,
starts at $0$, and increases towards $1$ over time (situations where it decreases are discussed later).
While the underlying population $P_0$ is known, the reach $\r$ is not
known and must be inferred from the data.

Substituting $P = \r P_0$, and integrating equation~\eqref{eq:NTa} over a day
gives a relationship that involves \textit{only measurable and computable quantities} $\T$, $\C_T = \T+\R_T$, and $\N_T$, and the parameters
of the model, namely
\be\label{eq:NTdisc}
\T(t) = \frac{1}{\bt} \N_T(t+1)
+ \frac{1}{\rt P_0} \C_T(t) \T(t) ,
\ee
where
\bd
\rt = \e \r (1-c) .
\ed
Note that $\N_T$ is shifted forward by one day since new infections reported on day $t+1$ are
determined by active infections and susceptible population on day $t$.
Eq.~(\ref{eq:NTdisc}) is the {\em fundamental equation} governing the pandemic.
It establishes a linear relationship between $\N_T$, $\T$, and $\C_T\T$,
that can be computed using the fundamental equation and the first three equations below (after fixing $\g$).

In addition to the fundamental equation, we will need discrete forms of other equations of the model to compute all quantities. We group them in two---first the quantities that can be computed from $\N_T$:
\begin{equation}\label{eq:Tdisc}
\begin{split}
\T(t) & = \N_T(t) + (1-\g) \T(t-1)  \\
\R_T(t) & = \R_T(t-1) + \g \T(t-1)  \\
\C_T(t) & = \T(t) + \R_T(t) = \N_T(t) + \C_T(t-1)
\end{split}
\end{equation}

The second group is of equations that involve numbers that cannot be computed from reported data:
\begin{equation} \label{eq:disc}
\begin{split}
\N(t) & =  \beta(1-\e) S(t-1)\M(t-1)  \\
\M(t) & = \N(t) + (1-\g)\M(t-1)  \\
\R(t) & = \R(t-1) + \g \M(t-1)  \\
\C(t) & = \M(t) + \R(t) = \N(t) + \C(t-1) \\
\U(t) & = \M(t) - \T(t)  \\
\R_U(t) & = \R(t) - \R_T(t) \\
S(t) & = 1 - \frac{\C(t)}{\r P_0}
\end{split}
\end{equation}
It is easy to see that all the quantities can be computed using above equations in addition to the fundamental equation once
the parameter values used in the equations are available.


\section{Fixing $\g$}\label{sec-gamma}

As discussed in the previous section, reported removal data does not provide a good estimate for $\g$.
In \cite{infection-duration}, median duration of infection for asymptomatic cases was estimated in the range $[6.5, 9.5]$
and mean duration for symptomatic cases in the range $[10.9, 15.8]$ days with a caveat that the duration reduces
when children are included. In~\cite{infection-duration-2}, infection duration for symptomatic cases was
observed to be less than $10$ days. Since our groups $U$ and $T$ consist of a mix of 
asymptomatic and symptomatic cases, and it is likely that an infected person stops infecting others before becoming
RTPCR negative, we take the mean duration of infection for both groups to be $10$ days, implying $\g = 0.1$.
All our simulations are done using the above value of $\g$ and show a good fit with the actual trajectories.

\section{Estimation of $\bt$ and $\rt$}\label{sec-detected-parameters}

One of the distinctive features of our approach is a methodology 
for estimating the values of all the parameters in the pandemic model from
reported raw data on the number of daily new cases.
The model has five parameters $\g$, $\beta$, $\e$, $\r$, and $c$. At a first glance, these appear all independent, however,
we show in section~\ref{sec-all-parameters} that last four are essentially determined by $\bt$ and $\rt$. In this section, we show how to
estimate $\bt$ and $\rt$ from reported data using the fundamental equation.

Let $\rN_T(t)$ be the reported new infections on day $t$. Note that $\rN_T(t)$ may not be the same as {\em detected} new infections on day $t$ since there may be delays in reporting detected cases. Moreover, weekends often see fewer tests being done, causing unexpected variations in $\rN_T$. To remove latter, 
we average $\rN_T(t)$ over a week, and let
\[ \aN_T(t) = \frac{1}{7}\sum_{j=0}^6 \rN_T(t-j). \]
Let $\aC_T(t) = \sum_{s=0}^t \aN_T(s)$, the total number of reported cases until day $t$, and $\aT(t)$ be the number
of reported active cases on day $t$ computed inductively using equation $\aT(t) = \aN_T(t) + (1-\g)\aT(t-1)$.

Fix a time interval $[t_0, t_1]$. Define ($t_1-t_0$)-dimensional vectors $\u$, $\v$, $\w$ as follows:
\bd
\u(t-t_0) = \aT(t), t_0 \leq t < t_1,
\ed
\bd
\v(t-t_0) = \aN_T(t+1), t_0 \leq t < t_1,
\ed
\bd
\w(t-t_0) = \frac{1}{P_0} \aC_T(t)\aT(t), t_0 \leq t < t_1.
\ed
Then the following \textit{linear regression problem} is solved:
\bd
\min_{\bt,\rt} || \u - \frac{1}{\bt} \v - \frac{1}{\rt } \w
||^2 .
\ed
The \textit{quality of the fit}
parameter, usually denoted by $R^2$, is computed as follows:
\bd
R^2 = 1 - \frac{ || \u - \frac{1}{\bt} \v - \frac{1}{\rt } \w ||^2 }
{ || \u ||^2 } ,
\ed
with the optimal parameter choices.
The closer $R^2$ is to one, the better is the quality of the fit.

At times, when there are relatively few data points ($t_1-t_0$ is small), or the data has significant errors, above linear regression method fails to work (e.g., estimated parameter value becomes negative).
In such situations we use a different method for estimation that is more tolerant to errors
as described below.

Let 
\begin{eqnarray*}
R^2_{\beta} & = & 1 - \frac{|\pvec{u}-\frac{1}{\bt}\pvec{v}-\frac{1}{\rt}\pvec{w}|^2}{|\pvec{u}-\frac{1}{\rt}\pvec{w}|^2} \\
R^2_{\r} & = & 1 - \frac{|\pvec{u}-\frac{1}{\bt}\pvec{v}-\frac{1}{\rt}\pvec{w}|^2}{|\pvec{u}-\frac{1}{\bt}\pvec{v}|^2}
\end{eqnarray*}
Find values of $\bt > 0$ and $\rt > 0$ that maximize the product $R^2 = R^2_{\beta}\cdot R^2_{\r}$. 
This choice ensures that both $\bt$ and $\rt$ play almost equally significant roles in minimizing the error.
Further, the desired maximum of $R^2_{\beta}R^2_{\r}$ is guaranteed to exist:

\begin{lemma}\label{l-alternate-R}
When $\pvec{u}$ is independent of $\pvec{v}$ as well as $\pvec{w}$, there is a maxima of $R^2$ with $R^2_{\beta},
R^2_{\r}, \bt, \rt > 0$.
\end{lemma}

The only situation when the above method will not yield the desired maxima of $R^2$ is when
$\pvec{u}$ is dependent on either $\pvec{v}$ or $\pvec{w}$. Former implies that $\T$ is proportional to
$\N_T$ over the time period, or equivalently, $S$ does not change over the period. This implies $\N = 0 = \N_T = \T$ for the
period. Similarly, latter implies that $\T$ is proportional to  $\C_T\T$ for the duration, or equivalently, $\C_T$
does not change over the period. This also implies that $\N_T = 0 = \N$. Either case occurs when the pandemic has
effectively ended and there are no new cases for an extended period.

The uncertainty in the parameter estimation is computed using the standard mean-square error formula for
linear regression. We use it to compute $95$\% confidence interval ranges for $\bt$ and $\rt$ values.

\section{Phases of the Pandemic}\label{sec-phases}

The parameters $\r$, $\beta$ and $\e$ are not constant, and vary over time. This causes changes in $\bt$ and $\rt$ as well.
The contact rate $\beta$ changes for following reasons:
\bit
\item Emergence of new and more infectious variants of the virus,
which would spread faster than its predecessor.
It takes time for the new variant to overtake whatever existed previously,
which is why this factor would cause $\beta$ to increase over a period.
\item Non-compliance with COVID guidelines.
The $\beta$ parameter measures the likelihood of infection when an
infected person (from either $U$ or $T$) meets a susceptible person from $S$.
Thus $\beta$ increases if people do not wear masks, or fail to maintain
social distancing, and the like.
\item The parameter can also \textit{decrease} suddenly, with almost a step change,
due to non-pharmaceutical interventions such as lockdowns.
\eit
The reach $\r$ changes for following reasons:
\bit
\item Spread of the pandemic to parts of the region that were previously untouched by it causes $\r$ to increase. The parts
may even be physically co-located with parts already touched by the pandemic comprising of those people who had completely isolated themselves.
\item Elimination of the pandemic from parts of the region that were under its influence causes $\r$ to decrease by the
fraction of still susceptible population of the parts.
\item Vaccination of susceptible people causes $\r$ to decrease, as these people moving out of susceptible compartment can
be viewed as effective population under the pandemic reducing. Similarly, loss of immunity among immune population 
causes $\r$ to increase as this can be viewed as effective
population under the pandemic increasing. This is formalized by the following lemma.
\eit

\begin{lemma}\label{l-gain-loss}
Suppose $\rgain$ is the fraction of susceptible population that became immune via vaccination, and $\rloss$ is the fraction of
immune population that lost immunity over a specified period of time. Then the new trajectory of the pandemic is obtained
by multiplying both $\beta$ and $\r$ (equivalently both $\bt$ and $\rt$) by $1+\frac{\rloss-\rgain}{\r}$.
\end{lemma}

Finally, the detection rate $\e$ may increase due to more comprehensive testing, and may decrease due to reduction in testing.

The changes in parameter values occur either as a slow drift over an extended period of time, or
as sudden rise and fall. We divide the entire timeline of the pandemic into \textit{phases},
such that within each phase, the parameters are (nearly) constant. A {\em phase change} occurs 
when one or more parameter values change significantly. It could be due to a quick change for reasons
listed above, or accumulated slow change over an extended period. By convention, we include the duration
of change in a parameter as part of new phase and call it {\em drift period} of the phase. The remaining
duration of a phase is called {\em stable period} of the phase.

When the value of $\e$ changes, then the relationship $T = \e M$ breaks down. The following lemma shows that 
$T$ converges to $\e M$ as soon as $\e$ stabilizes to its new value. 

\begin{lemma}\label{l-new-phase}
Suppose a new phase begins at time $t_0$ with a drift period of $d$ days. Further, suppose that the value of parameter $\e$ changes from $\e_0$ to $\e_1$ during the drift period. Then,
$\M(t_0+d) = \frac{1}{\e_1} \T(t_0+d)$.
\end{lemma}

The above analysis leads to the following methodology of phase identification and parameter estimation for phases:

\begin{enumerate}
\item
Suppose first phase starts at $t = 0$. Consider a small initial drift period $d$ (we start with $d=10$) and a small
time interval $[0, t_1]$, and compute the values of $\bt$ and $\rt$ for
this interval. 
\item
Increase the value of $t_1$ and adjust the value of $d$ until value of $R^2$ stabilizes. Freeze the computed values of
$\bt$ and $\rt$ for the phase.
\item
Increase the value of $t_1$ further until the fundamental equation has significant errors. This indicates that a new phase
has started.
\item
Repeat the same with every subsequent phase.
\end{enumerate}

We demonstrate the above methodology for one phase (phase $\#9$) in India: when the delta-variant started spreading rapidly in the country
during April 2021. In Appendix~\ref{sec-phase-plots}, we have plotted points $(\aT - \frac{1}{\bt} \aN_T, \frac{1}{P_0} \aC_T * \aT)$ for different values of $t_1$.
During the drift period of a phase, when the parameter values are changing, 
the points continuously drift away from a line passing through the origin (Figures~\ref{fig:IN-Ph-04-21}, \ref{fig:IN-Ph-04-23}, \ref{fig:IN-Ph-04-25}) indicating that equation~\ref{eq:NTdisc} is not satisfied. When the phase stabilizes, the points corresponding to the period line up nicely (Figures~\ref{fig:IN-Ph-04-27}, \ref{fig:IN-Ph-05-07}, \ref{fig:IN-Ph-05-17}, \ref{fig:IN-Ph-05-27}, \ref{fig:IN-Ph-06-07}) indicating that the equation~\ref{eq:NTdisc} is now satisfied.
The plots also show that values of $\bt$ and $\rt$ are changing quickly during the drift period, and do not change much during stable period. This leads to easy identification of phases and stable period within.

\subsection{Parameter values during drift period}

We have so far seen how to estimate values of $\bt$ and $\rt$ during stable
period of every phase.
However, in order to simulate the course of the pandemic,
it is necessary to have the values of the parameters
during the drift period as well.

Suppose $d$ is the number of days in drift period, and
$b_0$ and $b_1$ are the computed values of a parameter
in the previous and the current phases. Then its value will move from $b_0$ to $b_1$ during the drift period.
A natural way of fixing its value during the period is to use either arithmetic or geometric progression. That is,
on $i$th day in the drift period the value is set to $b_0 + \frac{i}{d}\cdot(b_1-b_0)$ or 
$b_0\cdot(\frac{b_1}{b_0})^{i/d}$ respectively.

Among these, geometric progression captures the way parameters change better:
\begin{itemize}
\item
When a new, more infectious, mutant spreads in a population, its infections grow exponentially initially. 
This corresponds to a multiplicative increase in $\beta$.
\item
Similarly, a new virus spreads in a region exponentially at the beginning. This corresponds to a multiplicative increase in
$\r$.
\item
A lockdown typically restricts movement sharply causing a multiplicative decrease in $\beta$.
\item
A change in testing strategy typically gets implement fast in a region, causing a multiplicative change in $\e$.
\end{itemize}
For these reasons, we assume that changes in parameters $\beta(1-\e)$ (this is the effective contact rate due to quarantining of detected cases), $\rho$, and $\e$ are multiplicative. Further, changes in parameter $c$ are additive as it is constant of integration ensuring continuity between two phases. Therefore, we may assume that changes in $1-c$ are multiplicative. This leads to the conclusion that changes in $\bt$ and $\rt$ are also multiplicative.

Having defined how the parameters change during drift periods, we assume that the equations
(\ref{eq:NTdisc}), (\ref{eq:Tdisc}), and (\ref{eq:disc}) hold {\em on all days}. When in drift period, even the parameter values in the
equations change daily as defined above. Subsequent sections show that our model with these assumptions is able to
capture the trajectory of the pandemic very well.

\section{Future Projections} \label{sec-future}

Once the quantities $\bt,\rt$ are estimated as above for current phase, equations (\ref{eq:NTdisc}) and (\ref{eq:Tdisc}) can be used to compute values of $\N_T$, $\C_T$, and $\T$ for the entire phase duration.
If the model captures the dynamics well, the predictions for daily new cases $\N_T$ should match closely with averaged reported numbers $\aN_T$ after the phase enters stable period as long as parameters do not change significantly. Indeed, this is confirmed by our simulations of trajectories in multiple countries. For example, for the phase $\#9$ of India discussed in the previous section, predicted trajectory changed rapidly when the phase was in drift period, and stabilized when it transitioned to stable period~(see Figure~\ref{fig:IN-Tr-9}).

\bfig[h!]
\bc
\includegraphics[height=70mm,width=160mm]{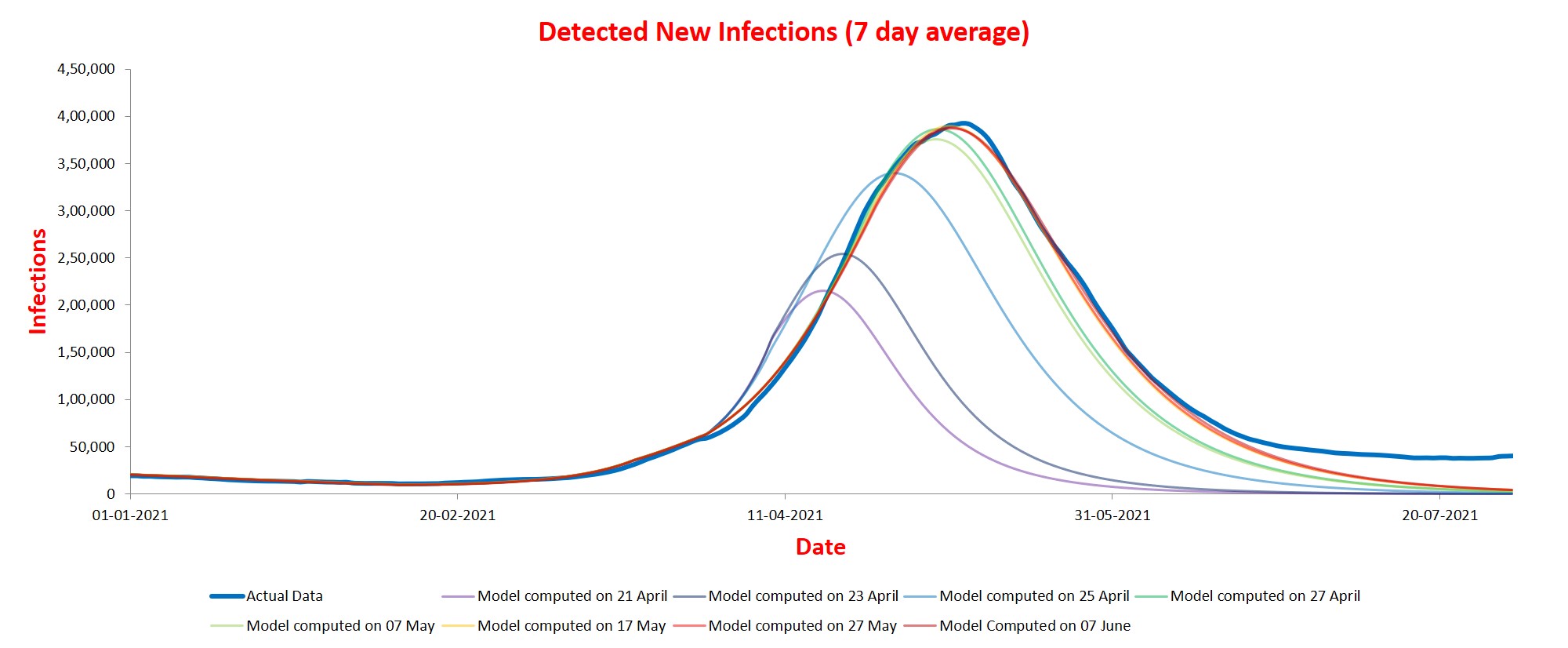}
\ec
\caption{Predicted Trajectories for India during April-June, 2021}
\label{fig:IN-Tr-9}
\efig

This property allows one to accurately predict the future course of the pandemic once the present phase stabilizes. We
used this to make several successful predictions in the past. 
Some notable ones were predicting the timing and height of the peak
of second wave of India ten days in advance~\cite{India-tweet-delta},
predicting timing of the peak of third wave in India as well as many states
of the country~\cite{India-tweet-omicron}, predicting timing and height of
the peak of Delta-wave in UK ten days in advance~\cite{UK-tweet-delta},
and predicting timing and height of the peak of Delta-wave in US more than
a month in advance~\cite{US-tweet-delta}.
The predictions for India and its states were useful to the policy-makers
in planning the required capacity for providing health care, and 
scheduling nonpharmaceutical interventions such as school reopenings.

\section{Estimation of $\r$ and $\e$}
\label{sec-all-parameters}

After fixing values of $\g$, the model has four parameters left: $\beta$, $\r$, $\e$ and $c$. 
We have seen how to estimate values of composite parameters $\bt$ and $\rt$ at all times,
which allows us to compute the trajectory of daily new detected cases for a region.
We can obtain more information about the pandemic if values of $\r$ and $\e$ can be
estimated separately. For example, $\e$ will enable us to estimate trajectory of total daily new cases, including undetected ones.
We provide more applications in the next two sections.

Without any additional information, besides the daily new detected infections time series, it is not possible to estimate
value of $\e$:

\begin{lemma}\label{l-any-eps}
Given detected new cases trajectory, $\N_T(t)$, $0 \leq t \leq t_F$, there exist infinitely many total new cases trajectories and  corresponding values of $\e$ consistent with $\N_T$.
\end{lemma}

In this section, we show that with {\em just one additional data point}---$\C(0)$ and $\M(0)$, the total number of cases up to time $t = 0$ and total active cases at $t=0$----the number of possible trajectories for $\N(t)$, consistent with the given data, becomes finite:

\begin{theorem}\label{t-finite-eps}
Given detected new cases trajectory, $\N_T(t)$, $0 \leq t \leq t_F$ and $\C(0)$, there exist only finitely many trajectories for $\N(t)$ consistent with $\N_T$. Further, a good estimate for all the trajectories can be obtained efficiently.
\end{theorem}

As is shown in the proof of above theorem (see Appendix~\ref{sec-proofs}), the trajectory for $\N(t)$ is unique for the first phase, but there may be multiple ones for subsequent phases, identified by a unique value for the pair $(\e, c)$ for each. Using the observation that the value of $\e$ from one phase to next will not change significantly, we can identify a unique {\em canonical} trajectory for total new cases: for each phase, given the possible values of $\e$ that give rise to consistent trajectories for $\N(t)$, choose the {\em canonical value} of $\e$ to be the one closest to the canonical value of $\e$ of previous phase (for first phase, there is anyway a unique value of $\e$). The corresponding trajectory for $\N(t)$ is
called {\em canonical trajectory}.

As the theorem also states, the canonical value of $\e$ and corresponding value of $c$ can be efficiently estimated. This, in turn, provides values of $\beta = \frac{\bt}{(1-\e)(1-c)}$ and $\r = \frac{\rt}{\e(1-c)}$ for the phase.

In this way, we get the values of all parameters at all times.

\subsection{Calibrating the Model}

Above shows how to estimate parameter values for all phases,
provided we know the values of $\C(0)$. This is equivalent to finding out the values of parameters $\e$ for the first phase, say $\e_1$, since $\C(0) = \frac{1}{\e_1} \C_T(0)$ (since $c_1 = 0$ as shown in the proof of theorem~\ref{t-finite-eps}).

While this is good in theory, we do not know $\e_1$ in practice. Moreover, time $t=0$ when the data becomes available for the first time is unlikely to be the time when the pandemic begins, and hence we may not have $c_1 = 0$.
However, $c_1 = 0$ will still be a good estimate since $R(0)$ is likely to be very small.
We estimate value of $\e_1$ from other information available about the pandemic. This is called \textit{calibrating the model}.
We can calibrate the model in two ways:

\bit
\item
A sero-survey at time $t_0$ provides a good estimate of $\C(t_0-\delta)$, where $\delta$ equals the
time taken for antibodies to develop. 
Once we accurately estimate $\e_1$, the model can compute $\C(t)$
at all times $t$.
We choose a suitable value of $\e_1$ ensuring that model computation
matches with the sero-survey result at time $t-\delta$.
\item
When the pandemic has been active long enough in a region without major,
long-term restrictions, we may assume that it has reached all sections
of society, making $\r$ close to $1$.
Again, we can choose $\e_1$ that ensures that the reach of the pandemic
is close to $1$ at suitable time.
\eit

While using the above two methods for calibrating the model, following points need to be kept in mind:

\begin{description}
\item[Using serosurveys.]
Many serosurveys suffer from significant sampling biases. For example, if a survey is done using residual sera from a period of high infection numbers, it is likely to significantly overestimate the seroprevalence because a large fraction of uninfected persons would not venture to give blood sample in such a period.
In order to minimize sampling biases, therefore, one should use serosurveys done during a period of low infection numbers. Even then, some uninfected people may not participate making the estimates higher than actual. To further reduce bias, one should ideally be able to use multiple serosurveys as well as use the fact that reach is close to $1$ by a given time.

\item[Using reach.]
As observed earlier (Lemma~\ref{l-gain-loss}), parameter $\r$ is impacted by several factors, including gain and loss of immunity.
Therefore, $\r$ may not be close to $1$ even when the pandemic has spread over entire population.
To capture this, we define $\ra$ to denote the actual reach of pandemic, so $\r = \ra \cdot (1 + \frac{\rloss-\rgain}{\ra}) = \ra + \rloss - \rgain$. To calibrate the model using $\ra \approx 1$ at certain time, we need an estimate of $\rloss -\rgain$ too, which
introduces more errors in calibration.
\end{description}

There are regions where neither an accurate sero-survey is
available, and it is evident that reach is nowhere close to $1$.
For such regions, calibration cannot be done with any confidence,
and so estimation of all parameter values is not possible.

\section{Analysis of the Past}
\label{sec-past}

The parameter table of a country enables us to quantify the impacts of
various events like the arrival of a new mutant, or a lockdown.
Moreover, through the reach parameter, we can also explain the somewhat
mysterious phenomenon of multiple peaks occurring in rapid succession
that was observed in many countries.

Below, we discuss in detail the progression of the pandemic in four countries. 
The time series data for India was sourced from~\cite{covid19India}, and for rest of the countries from~\cite{worldometers}.

\subsection{India}

Model computed trajectory for detected cases shows an excellent match with the reported trajectory:

\bfig[h!]
\bc
\includegraphics[width=160mm]{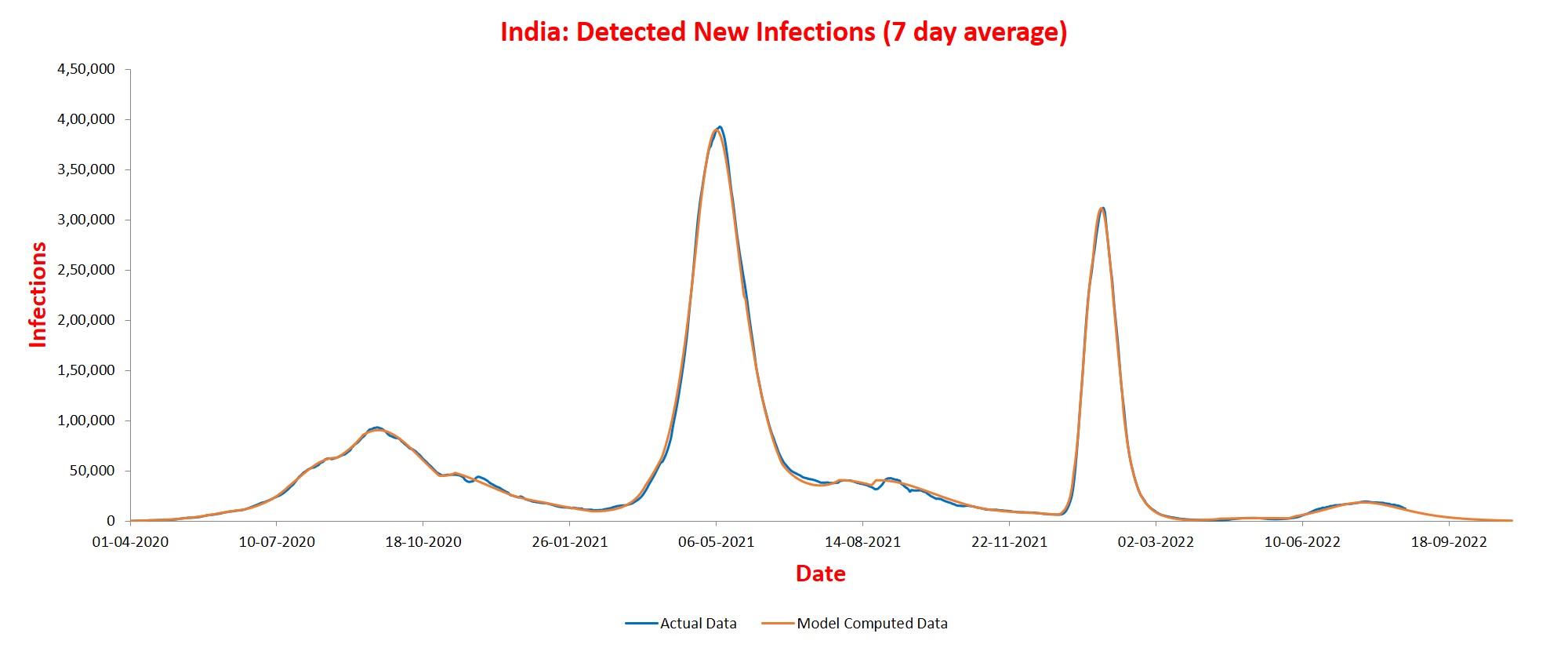}
\ec
\caption{Predicted and Actual Trajectories for India}
\label{fig:IN}
\efig

For estimating all parameters, the calibration was done using sero survey done in December
2020~\cite{India-serosurvey-three}, a period of low infection.
Estimates of seropositivity computed by the model were matched with two
other serosurveys~\cite{India-serosurvey-two,India-serosurvey-four},
and very good agreement was found.
Further, our model showed that the reach was close to maximum by December 2021,
a very likely scenario.
Interestingly, the detection rate $\e$ stayed almost unchanged at $1/32$
throughout the  course of the pandemic.
Comparing the timeline of the pandemic~\cite{timeline-India} with parameter table below, we observe the following.

\begin{table}[h!]
\caption{Parameter Table for India}
\bc
\begin{tabular}{|c|c|c|c|c|c|}
\hline
Ph No & Start & Drift & $\beta$ & $1/\e$ & $\r$ \\
\hline\hline
1 & 03-03-2020 & 4 & $0.29 \pm 0.04$ & {\color{red} $32$} & $0 \pm 0$ \\ 
\hline
2 & 19-03-2020 & 1 & $0.33 \pm 0.02$ & $32 \pm 0$ & $0 \pm 0$ \\ 
\hline
3 & 12-04-2020 & 4 & $0.16 \pm 0$ & $32 \pm 0$ & $0.033 \pm 0.003$ \\ 
\hline
4 & 17-06-2020 & 30 & $0.16 \pm 0.01$ & $32 \pm 0$ & $0.204 \pm 0.02$ \\ 
\hline
5 & 20-08-2020 & 17 & $0.16 \pm 0$ & $32 \pm 0$ & $0.364 \pm 0.012$ \\ 
\hline
6 & 29-10-2020 & 10 & $0.18 \pm 0$ & $32 \pm 0$ & $0.423 \pm 0.008$ \\ 
\hline
7 & 18-12-2020 & 20 & $0.19 \pm 0.01$ & $32 \pm 0$ & $0.448 \pm 0.005$ \\ 
\hline
8 & 11-02-2021 & 35 & $0.38 \pm 0.01$ & $32 \pm 0$ & $0.462 \pm 0.008$ \\ 
\hline
9 & 30-03-2021 & 25 & $0.28 \pm 0.01$ & $32 \pm 0$ & $0.828 \pm 0.009$ \\ 
\hline
10 & 25-05-2021 & 0 & $0.27 \pm 0$ & $32 \pm 0$ & $0.859 \pm 0.026$ \\ 
\hline
11 & 20-06-2021 & 38 & $0.51 \pm 0.01$ & $32 \pm 0$ & $0.92 \pm 0.001$ \\ 
\hline
12 & 20-08-2021 & 2 & $0.58 \pm 0.02$ & $32 \pm 0$ & $0.92 \pm 0.004$ \\ 
\hline
13 & 01-11-2021 & 35 & $0.61 \pm 0.01$ & $32 \pm 0$ & $0.949 \pm 0.001$ \\ 
\hline
14 & 26-12-2021 & 9 & $1.56 \pm 0.21$ & $32 \pm 0$ & $1.03 \pm 0.031$ \\ 
\hline
15 & 10-01-2022 & 7 & $1.18 \pm 0.02$ & $32.1 \pm 0.1$ & $1.035 \pm 0.022$ \\ 
\hline
16 & 06-02-2022 & 1 & $1.56 \pm 0.01$ & $32.1 \pm 0$ & $1.02 \pm 0.011$ \\ 
\hline
17 & 24-03-2022 & 20 & $3.45 \pm 0.26$ & $32.1 \pm 0$ & $1.044 \pm 0.003$ \\ 
\hline
18 & 02-06-2022 & 5 & $2.89 \pm 0.06$ & $32.4 \pm 2.1$ & $1.078 \pm 0.113$ \\ 
\hline
 \end{tabular}
\ec
\label{table:India}
\end{table}

\begin{description}
\item[First wave (March to October 2020):] The strict lockdown imposed at the
end  of March 2020 brought down the contact rate $\beta$ by a factor of two.
The reach was very small until May ($\approx 0.03$) but increased to $0.36$ 
between the end of June and the end of August.
This was caused by reverse migration of workers and a partial lifting of lockdown that
happened during this period.
\item[Second wave (February to July 2021):] The arrival of the Delta variant
caused the value of $\beta$ to rise to $0.38$ in February 2021.
As the variant began to spread in different parts of the country,
most states imposed restrictions, which reduced the nationwide
$\beta$ to $0.28$ by April.
In the same month, $\r$ increased sharply to $0.83$. 

Note that while increase in $\r$ was clearly due to Delta variant, the increase happened more than a month after increase in $\beta$. We have observed this delayed increase phenomenon in $\r$ repeatedly.

The removal of all restrictions by August caused $\beta$ to increase to $0.58$.
This suggests that the Delta variant was more infectious by a factor of
$\approx 2$ compared to original variant.
\item[Third wave (December 2021 to March 2022):]
The arrival of the Omicron variant caused $\beta$ to increase sharply to
$1.56$ and $\r$ to increase to $1.03$ (from $0.95$) by the end of December.
In January, mild restrictions were imposed across the country, causing
$\beta$ to drop to $1.18$.
These were lifted in February, and $\beta$ went back up to $1.56$ in February.

\item[A ripple (April 2022 to September 2022):] The value of $\beta$ increased to around $3$ by June. This, coupled with an increase in $\r$ from $1.02$ to $1.08$ (indicating around $6\%$ population losing natural immunity) caused a ripple
that peaked in July.
\end{description}

At present, around $98\%$ of population is estimated to have natural immunity.

\newpage

\subsection{UK}

Model computed trajectory for detected cases shows a good match with the reported trajectory:

\bfig[h!]
\bc
\includegraphics[width=160mm]{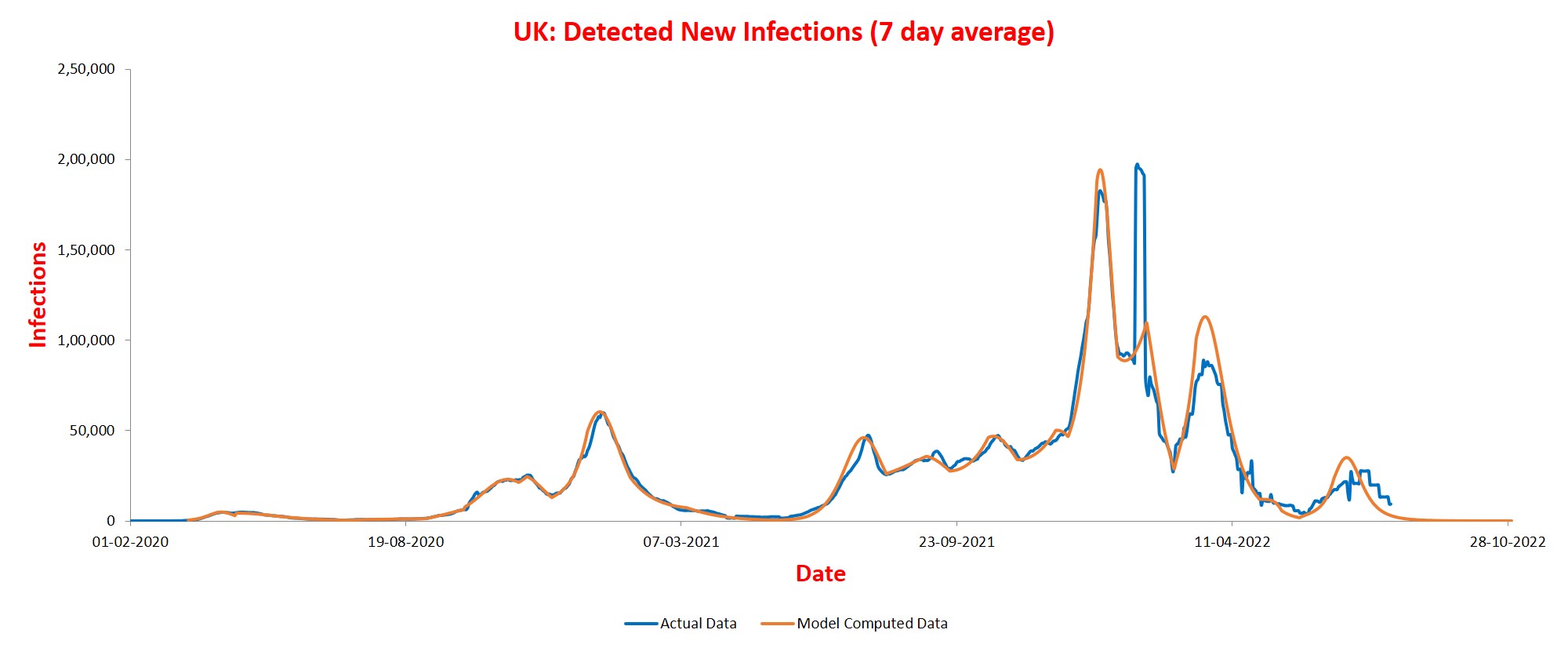}
\ec
\caption{Predicted and Actual Trajectories for UK}
\label{fig:UK}
\efig

For estimating all parameters, the
calibration was done using two of the three serosurveys reported
in~\cite{UK-serosurvey}.
During the period of serosurveys, the infection numbers were going up and down,
which made calibration a little tricky.
We used the numbers from the last two surveys as well as the observation that reach has remained stationary from November 2021 (suggesting that $\ra$ 
has been around $1$ since then) for calibration. The detection rate started at $1/9.3$ and over time increased to almost one in three cases.
Comparing the timeline of the pandemic~\cite{timeline-UK} with parameter table below, we observe the following.

\begin{table}[h!]
\caption{Parameter Table for UK}
\bc
\begin{tabular}{|c|c|c|c|c|c|}
\hline
Ph No & Start & Drift & $\beta$ & $1/\e$ & $\r$ \\
\hline\hline
1 & 14-03-2020 & 10 & $0.26 \pm 0.01$ & {\color{red} $9.3$} & $0.02 \pm 0.001$ \\ 
\hline
2 & 17-04-2020 & 0 & $0.15 \pm 0$ & $9.3 \pm 0$ & $0.063 \pm 0.003$ \\ 
\hline
3 & 07-07-2020 & 10 & $0.24 \pm 0.01$ & $9.3 \pm 0$ & $0.08 \pm 0.002$ \\ 
\hline
4 & 01-09-2020 & 7 & $0.27 \pm 0.01$ & $9.3 \pm 0$ & $0.122 \pm 0.004$ \\ 
\hline
5 & 30-09-2020 & 0 & $0.21 \pm 0.01$ & $9.3 \pm 0$ & $0.296 \pm 0.032$ \\ 
\hline
6 & 09-11-2020 & 5 & $0.26 \pm 0.01$ & $9.3 \pm 0$ & $0.3 \pm 0.021$ \\ 
\hline
7 & 03-12-2020 & 25 & $0.32 \pm 0.01$ & $8.7 \pm 1.2$ & $0.594 \pm 0.087$ \\ 
\hline
8 & 29-01-2021 & 40 & $0.68 \pm 0.04$ & $8.7 \pm 0.3$ & $0.616 \pm 0.053$ \\ 
\hline
9 & 15-05-2021 & 25 & $0.61 \pm 0.03$ & $7.7 \pm 1.9$ & $0.763 \pm 0.116$ \\ 
\hline
10 & 03-08-2021 & 28 & $0.37 \pm 0.01$ & $6.5 \pm 0.1$ & $0.921 \pm 0.043$ \\ 
\hline
11 & 18-09-2021 & 27 & $0.5 \pm 0.02$ & $5.9 \pm 0.2$ & $0.956 \pm 0.032$ \\ 
\hline
12 & 06-11-2021 & 27 & $0.54 \pm 0.01$ & $5.4 \pm 0.1$ & $1.046 \pm 0.034$ \\ 
\hline
13 & 13-12-2021 & 20 & $0.74 \pm 0$ & $4.2 \pm 0$ & $1.014 \pm 0.001$ \\ 
\hline
14 & 18-01-2022 & 20 & $0.87 \pm 0.19$ & $3.5 \pm 0.2$ & $1.013 \pm 0.035$ \\ 
\hline
15 & 28-02-2022 & 15 & $0.93 \pm 0.03$ & $3.1 \pm 0.1$ & $1.048 \pm 0.007$ \\ 
\hline
16 & 14-04-2022 & 10 & $0.73 \pm 0.09$ & $2.9 \pm 1.6$ & $1.019 \pm 0.343$ \\ 
\hline
17 & 01-06-2022 & 3 & $1.71 \pm 0.09$ & $2.8 \pm 0.4$ & $1.02 \pm 0.106$ \\ 
\hline
 \end{tabular}
\ec
\label{table:UK}
\end{table}

\begin{description}
\item[First wave (March to July 2020):] The strict lockdown imposed in March 2020 brought down the contact rate $\beta$ from $0.26$ to $0.15$ in mid-April. However, almost simultaneously, $\r$ increased three-fold causing another peak. By July, $\beta$ was back up to $0.24$ after removal of restrictions.

\item[Second wave (September 2020 to January 2021):] This wave was primarily caused by increase in value of $\r$ from $0.08$ to $0.6$. This increase in $\r$ was a natural consequence of very small effective population until August (less than one percent) and easing of lockdown from July (as noted above, increase in $\r$ happens with a lag). As the numbers started increasing, fresh restrictions were put in place bringing $\beta$ down by $20\%$ by September-end. This caused the cases to peak by October-end (by that time $\r$ increased to $0.3$). As the lockdown was eased before Christmas, both $\beta$ and $\r$ started increasing causing and second bigger peak in January 2021.

The rise in case number caused another lockdown, but this time $\beta$ did not decrease. Note that a new variant, called Alpha, started spreading in UK rapidly in December 2020. It was believed to be significantly more infectious than earlier one. This appears to be the reason why the value of $\beta$ did not decrease in January, and went up slightly instead.

\item[Third wave (February 2021 to October 2021):]
Lockdown was eased during February-March which resulted in a significant rise in $\beta$ to $0.68$ by second half of March. This jump, however, caused only a slight change in trajectory because reach stayed around
$0.6$ and more than $85\%$ of population within reach had natural immunity by then. Numbers started rising from mid-June due to increase in $\r$ (again a delayed increase).
There were three peaks in quick succession:
The first caused by increase in $\r$ to $0.76$ in July, 
the second caused by further increase in $\r$ to $0.92$ in August 
(when $\beta$ came down to $0.37$ during this period, likely caused by
precautions taken by people due to high numbers), and the
third caused by increase in $\beta$ to $0.5$ in addition to a slight increase in $\r$. This increase in $\beta$ was likely due to the Delta variant now active in the country.

\item[Fourth wave (November 2021 to August 2022):] In November the Omicron variant arrived causing $\beta$ to increase further. The wave had four peaks (although the second one got a bit messed up due to reporting of very large numbers on 31st January of backlog cases). These peaks were all caused by increase in $\beta$ -- to $0.74$ in December, to $0.87$ in January-end, to $0.93$ in March, and finally to $1.71$ in June. The stepwise increase is connected to levels of restrictions imposed.
\end{description}

At present, around $92\%$ of population is estimated to have natural immunity.

\newpage

\subsection{US}

Model computed trajectory for detected cases shows an excellent match with the reported trajectory:

\bfig[h!]
\bc
\includegraphics[height=80mm]{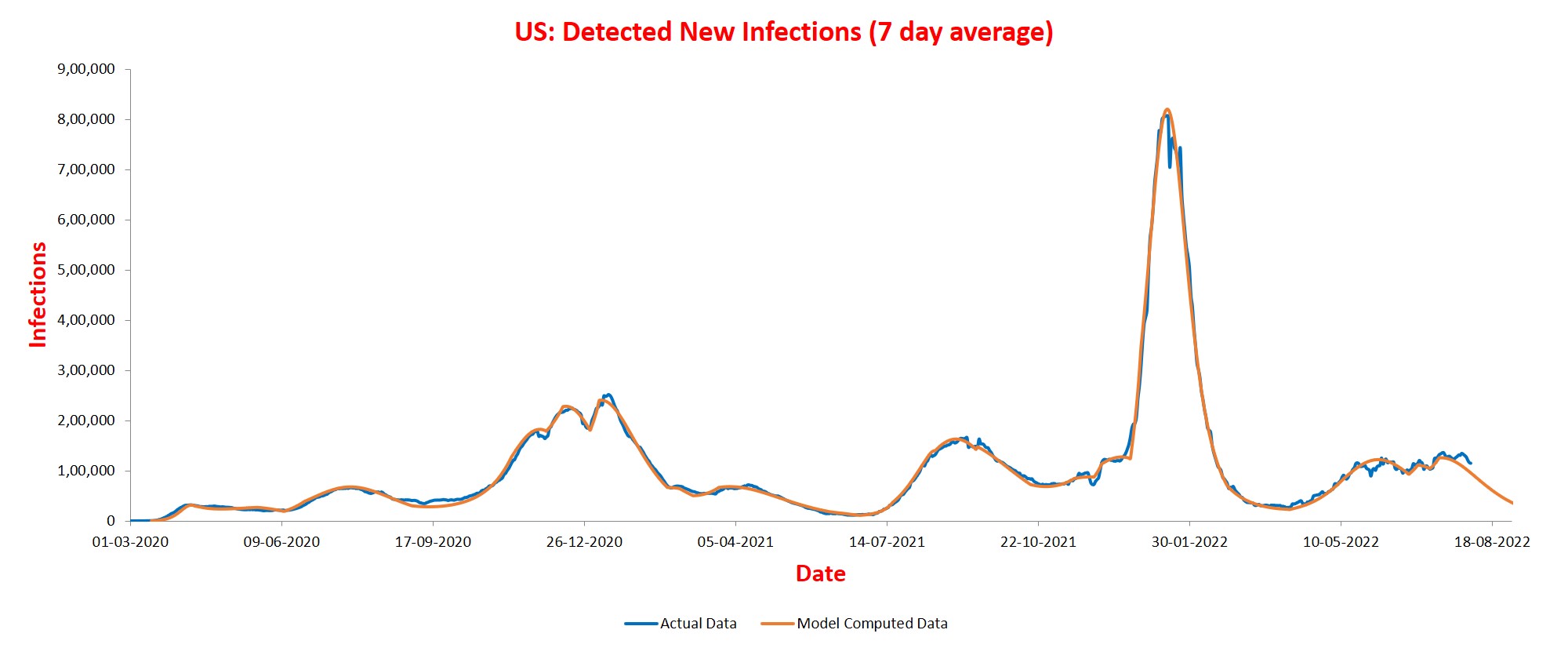}
\ec
\caption{Predicted and Actual Trajectories for USA}
\label{fig:US}
\efig

For estimating all parameters, the
calibration was done using the serosurvey~\cite{US-serosurvey}.
The samples were taken from life insurance applications.
The calibration was further supported by the fact that $\r$ has not
changed since December, suggesting that $\ra$ was close to $1$ at the time.
The detection rate has slowly decreased from $1/3.5$ to $1/4$ during the
course of the pandemic.
Comparing the timeline of the pandemic~\cite{timeline-US} with parameter table below, we observe the following.

\begin{table}[h!]
\caption{Parameter Table for US}
\bc
\begin{tabular}{|c|c|c|c|c|c|}
\hline
Ph No & Start & Drift & $\beta$ & $1/\e$ & $\r$ \\
\hline\hline
1 & 15-03-2020 & 3 & $0.31 \pm 0.02$ & {\color{red} $3.5$} & $0.007 \pm 0.001$ \\ 
\hline
2 & 13-04-2020 & 40 & $0.18 \pm 0.01$ & $3.5 \pm 0$ & $0.038 \pm 0.003$ \\ 
\hline
3 & 11-06-2020 & 12 & $0.18 \pm 0.01$ & $3.5 \pm 0$ & $0.113 \pm 0.006$ \\ 
\hline
4 & 03-09-2020 & 65 & $0.24 \pm 0.01$ & $3.5 \pm 0$ & $0.255 \pm 0.019$ \\ 
\hline
5 & 01-12-2020 & 10 & $0.24 \pm 0$ & $3.5 \pm 0$ & $0.325 \pm 0.007$ \\ 
\hline
6 & 30-12-2020 & 5 & $0.26 \pm 0.01$ & $3.5 \pm 0$ & $0.391 \pm 0.022$ \\ 
\hline
7 & 19-02-2021 & 7 & $0.23 \pm 0.01$ & $3.5 \pm 0$ & $0.462 \pm 0.009$ \\ 
\hline
8 & 08-03-2021 & 16 & $0.45 \pm 0.02$ & $3.6 \pm 0.6$ & $0.434 \pm 0.12$ \\ 
\hline
9 & 06-06-2021 & 10 & $0.38 \pm 0.01$ & $3.6 \pm 0$ & $0.459 \pm 0.002$ \\ 
\hline
10 & 26-06-2021 & 21 & $0.65 \pm 0.01$ & $3.7 \pm 0.1$ & $0.512 \pm 0.044$ \\ 
\hline
11 & 11-08-2021 & 3 & $0.46 \pm 0.01$ & $3.8 \pm 0.1$ & $0.583 \pm 0.036$ \\ 
\hline
12 & 11-09-2021 & 0 & $0.31 \pm 0.01$ & $3.8 \pm 0.1$ & $0.697 \pm 0.09$ \\ 
\hline
13 & 17-10-2021 & 28 & $0.42 \pm 0.01$ & $3.9 \pm 0$ & $0.756 \pm 0.007$ \\ 
\hline
14 & 28-11-2021 & 4 & $0.6 \pm 0.01$ & $3.9 \pm 0$ & $0.734 \pm 0.003$ \\ 
\hline
15 & 22-12-2021 & 6 & $0.53 \pm 0.01$ & $4.2 \pm 0.2$ & $1.088 \pm 0.038$ \\ 
\hline
16 & 24-02-2022 & 39 & $1.87 \pm 0.02$ & $4.3 \pm 0$ & $1.083 \pm 0.026$ \\ 
\hline
17 & 06-04-2022 & 35 & $1.02 \pm 0.02$ & $4.2 \pm 0.2$ & $1.195 \pm 0.032$ \\ 
\hline
18 & 24-06-2022 & 5 & $1.03 \pm 0.68$ & $4.1 \pm 0$ & $1.201 \pm 0.056$ \\ 
\hline
19 & 08-07-2022 & 5 & $0.71 \pm 0.07$ & $4 \pm 0$ & $1.279 \pm 0.018$ \\
\hline
19 & 08-07-2022 & 5 & $0.58 \pm 0.06$ & $4 \pm 0$ & $1.317 \pm 0.024$ \\ 
\hline
 \end{tabular}
\ec
\label{table:US}
\end{table}

\begin{description}
\item[First wave (March to August 2020):] Restrictions imposed in April 2020 brought down the contact rate $\beta$ from
$0.31$ to $0.18$ by mid-May. However, almost simultaneously, $\r$ increased to $0.04$ causing a flat trajectory. In June, most restrictions were lifted. This increased $\r$ further to $0.11$ causing a peak in July-end. The value of $\beta$, however, did not increase. This could be due to precautions taken by a large number of people.

\item[Second wave (September 2020 to February 2021):] By October, $\beta$ went up to $0.24$ and stayed around this value until the end of the wave. The value of $\r$ increased in three steps: to $0.26$ during September-October period, to $0.33$ in December, and to $0.39$ in January. This causes three successive peaks in November, December, and January.

\item[Third wave (March 2021 to November 2021):] There were two peaks separated
by more than four months in this period.
The Delta variant appeared to have arrived in March causing $\beta$ to
increase to $0.45$. However, it caused only a small peak since $\r$ stayed around $0.5$ until July, and more than $75\%$ of population under reach had natural immunity.
The reach started increasing in August to eventually become $0.7$ by mid-September causing another peak (yet another case of delayed increase in $\r$).

\item[Fourth wave (December 2021 to March 2022):]  The Omicron variant started
spreading in December causing $\beta$ to increase to $0.6$, but the numbers did not increase much by December-end,
since $\r$ did not change by much.
Then the reach increased substantially to $1.08$ in a short time leading to a very sharp and high peak. By February, the wave subsided, and even though $\beta$ jumped to $1.87$ in March, it did not cause cases to increase as more than $90\%$ of population was immune by then.

\item[Fifth wave (April 2022 to September 2022):] The primary cause of this wave appears to be a {\em loss of natural immunity}. By May, $\r$ increased to $1.2$ and is close to $1.3$ at present. This implies that more than $20\%$ of population has lost natural immunity in past six months. Immunity loss at such a scale has not been observed in the other three countries discussed here. Reasons for this are not clear. 
\end{description}

At present, around $85\%$ of population is estimated to have natural immunity.

\newpage

\subsection{South Africa}

Model computed trajectory for detected cases shows an excellent match with the reported trajectory:

\bfig[h!]
\bc
\includegraphics[height=80mm]{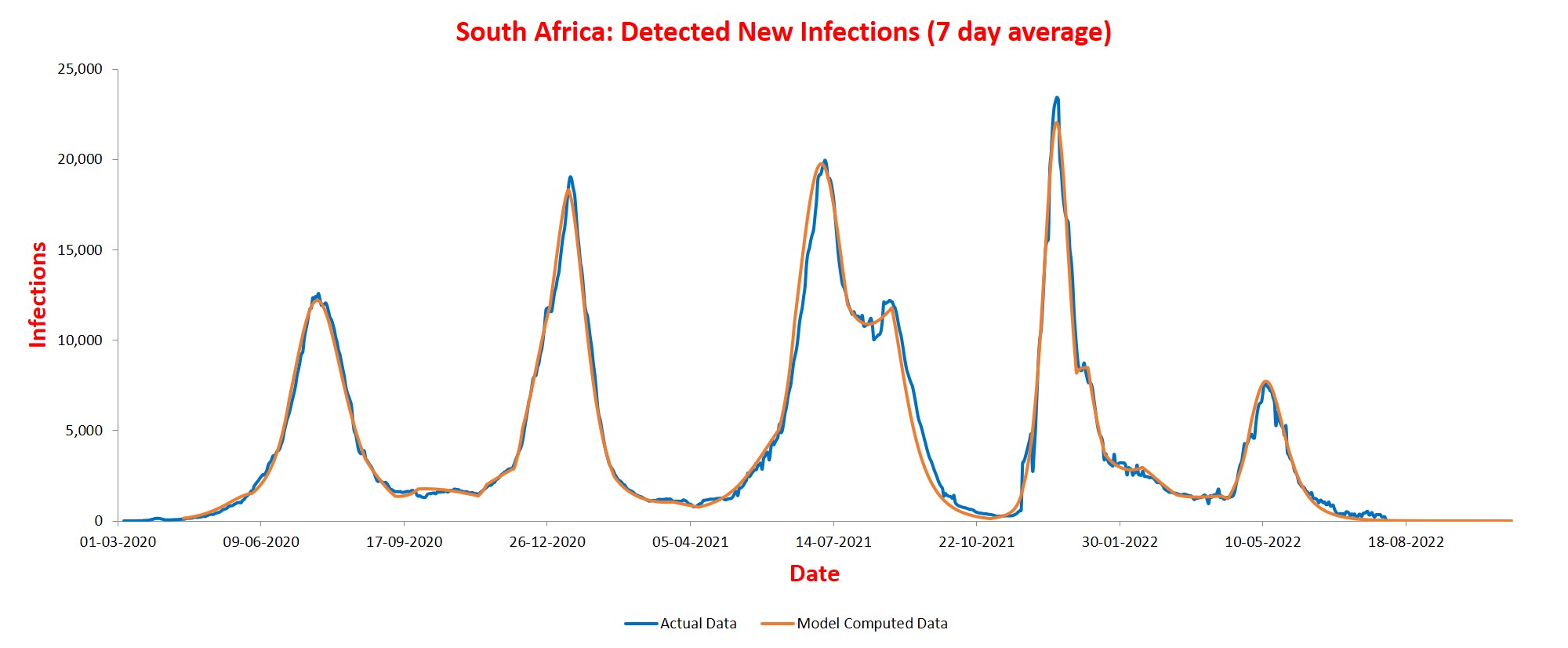}
\ec
\caption{Predicted and Actual Trajectories for South Africa}
\label{fig:SA}
\efig

For estimating all parameters, the calibration was done using the serosurvey~\cite{SA-serosurvey}.
The calibration was further supported by the fact that $\r$ has not changed since November suggesting that $\ra$ has been close to $1$ since then. The detection rate has remained almost unchanged at $1/17$ during the course of the pandemic.
Comparing the timeline of the pandemic~\cite{timeline-SA} with parameter table below, we observe the following.

\begin{table}[h!]
\caption{Parameter Table for South Africa}
\bc
\begin{tabular}{|c|c|c|c|c|c|}
\hline
Ph No & Start & Drift & $\beta$ & $1/\e$ & $\r$ \\
\hline\hline
1 & 16-04-2020 & 15 & $0.18 \pm 0.01$ & {\color{red} $17$} & $0.036 \pm 0.009$ \\ 
\hline
2 & 03-06-2020 & 25 & $0.19 \pm 0.01$ & $17.1 \pm 0$ & $0.241 \pm 0.005$ \\ 
\hline
3 & 21-08-2020 & 10 & $0.16 \pm 0.01$ & $17.1 \pm 0$ & $0.288 \pm 0.007$ \\ 
\hline
4 & 11-09-2020 & 15 & $0.25 \pm 0.01$ & $17.1 \pm 0$ & $0.329 \pm 0.007$ \\ 
\hline
5 & 08-11-2020 & 5 & $0.27 \pm 0.01$ & $17.1 \pm 0$ & $0.4 \pm 0.007$ \\ 
\hline
6 & 03-12-2020 & 5 & $0.31 \pm 0$ & $17.1 \pm 0$ & $0.5 \pm 0.037$ \\ 
\hline
7 & 28-12-2020 & 12 & $0.46 \pm 0.01$ & $17.9 \pm 0.1$ & $0.485 \pm 0.009$ \\ 
\hline
8 & 11-02-2021 & 40 & $0.65 \pm 0.02$ & $18 \pm 0$ & $0.536 \pm 0.002$ \\ 
\hline
9 & 11-04-2021 & 36 & $0.4 \pm 0.01$ & $18.1 \pm 0$ & $0.752 \pm 0.004$ \\ 
\hline
10 & 06-06-2021 & 10 & $0.43 \pm 0.01$ & $19.6 \pm 0.4$ & $0.938 \pm 0.037$ \\ 
\hline
11 & 24-07-2021 & 30 & $0.99 \pm 0.02$& $18.2 \pm 2.1$ & $0.923 \pm 0.108$ \\ 
\hline
12 & 01-11-2021 & 22 & $1.58 \pm 0.05$ & $17.5 \pm 1.1$ & $1.033 \pm 0.026$ \\ 
\hline
13 & 31-12-2021 & 7 & $1.27 \pm 0.01$ & $16.4 \pm 0.1$ & $1.01 \pm 0.01$ \\ 
\hline
14 & 20-01-2022 & 25 & $1.56 \pm 0.01$ & $16.2 \pm 0$ & $1.044 \pm 0.001$ \\ 
\hline
15 & 11-03-2022 & 30 & $4.35 \pm 0.04$ & $16.1 \pm 0$ & $1.027 \pm 0$ \\ 
\hline
16 & 16-04-2022 & 15 & $2.99 \pm 0.02$ & $15.8 \pm 0.2$ & $1.06 \pm 0.004$ \\ 
\hline
17 & 12-06-2022 & 38 & $2.56 \pm 0.15$ & $15.6 \pm 0$ & $1.076 \pm 0.002$ \\ 
\hline
 \end{tabular}
\ec
\label{table:SA}
\end{table}

\begin{description}
\item[First wave (April to August 2020):] Restrictions imposed in March and April 2020 brought down the contact rate $\beta$ to around $0.2$. A significant increase in $\r$ to $0.24$ by June-end caused the first wave that peaked in July-end.

\item[Second wave (September 2020 to February 2021):] Restrictions were lowered in September causing increase in $\beta$ value to $0.25$. Reach also continued to increase slowly to $0.5$. This caused only a slow rise since more than $40\%$ of population under the reach was already immune. The Beta variant arrived in December causing an immediate jump in $\beta$ value to $0.46$. This caused the second peak in January. Restrictions were reimposed in December to control the rise in the numbers due to Beta variant. The fact that $\beta$ still increased substantially shows that infectiousness of this variant was quite high.

\item[Third wave (March 2021 to October 2021):] With removal of restrictions measures by March, value of $\beta$ further increased to $0.65$. However, since increase in $\r$ is typically delayed and more than $90\%$ of population within reach was already immune, the rise in $\beta$ did not cause increase in numbers. The numbers started rising when $\r$ started increasing in April to become $0.75$ by mid-May. Restrictions were partly brought back causing $\beta$ to come down to around $0.4$. The value of $\r$ further went up to $0.94$ causing a peak in July.

As the numbers started coming down from the peak in August, an unusual phenomenon occurred. Cases started increasing once again, there was a short peak in second half of August, and then the numbers came down once again but with a slightly less steep slope than before. Our model shows that this happened due to a sharp increase in value of $\beta$ to nearly $1$ from $0.43$. Note that restrictions were being increased during June-July and were relaxed only from September, so the increase in $\beta$ was not due to relaxations. Was this caused by Omicron variant that was detected later in South Africa? The sharp increase in $\beta$ which then stayed high certainly suggests so.

\item[Fourth wave (November 2021 to March 2022):] 
In November, with more and more relaxations, Omicron caused $\beta$ to further increase to $1.58$ and
$\r$ to $1.03$ resulting in a high peak. No restrictions were imposed this time, and so the numbers rose and fell sharply.

\item[Fifth wave (March 2022 to June 2022):] A further increase in $\beta$ to around $3$ by April resulted in another peak in mid-May. This peak was, however, a small one since reach was stationary around $1.05$ and more than $95\%$ of population had natural immunity.
\end{description}

At present, around $98\%$ of population is estimated to have natural immunity.

\section{Analysis of the Immunity Loss}\label{sec-immunity}

After the South African authorities announced the emergence of a new
variant of concern (VOC), later named Omicron, the epidemiology community
started analysing the ability of the Omicron variant to bypass immunity
provided by vaccination, or prior exposure, or both.
Our objective in this section is to provide a \textit{quantitative} analysis
using the SUTRA model.
But before that, we give a brief summary of the vast literature 
based on laboratory (as opposed to population-level) studies.

Everywhere in the world where it was discovered, the Omicron VOC soon
replaced all other variants and was responsible for a massive increase
in cases.
This was due to high transmissibility conferred by the
mutation, ensuring a tight binding to the ACE 2 receptor facilitating
immune escape \cite{Denji-et-al22}.
The immune escape phenomenon was reported by many groups studying the
neutralization activity of sera from both infected and vaccinated individuals;
see \cite{Zhang-et-al22,Ren-et-al22,Cele-et-al21,Zeng-et-al21,Ai-et-al22}.
The immunity conferred by complete vaccination decreased from
80\% for the Delta variant to about 30\% for the Omicron variant.
People 
infected with the Delta 
were better off than those infected with the initial Beta variant.
There was a complete loss of neutralizing antibodies in over 50\% of
the vaccinated individuals and the decrease in titres varied from 43-122 fold
between vaccines \cite{Garcia-et-al22}.
A booster Pfizer dose could generate an anti-Omicron neutralizing response,
but titres were 6-23 fold lower than those for Delta variant.
Sera from vaccinated individual of the Pfizer or Astra Zeneca vaccine barely
inhibited the Omicron variant  five months after complete vaccination
\cite{Planas-et-al22}.
In addition, Omicron was completely or partially resistant to
neutralization by all monoclonal antibodies tested \cite{Denji-et-al22}.
Overall, most studies confirmed that sera from convalescent as well as
fully vaccinated individuals irrespective of the vaccine
(BNT162b2, mRNA-1273, Ad26.COV2.5 or ChAdOx1-nCoV19, Sputnik V or BBIBP-CorV)
contained very low to undetectable levels of nAbs against Omicron.
A booster with a third dose of mRNA vaccine appeared to restore neutralizing
activity but the duration over which this effect may last has not been
confirmed.
Double vaccination followed by Delta breakthrough infection, or prior infection
followed by mRNA vaccine double vaccination, appear to generate increased
protective levels of neutralizing antibodies \cite{Fleming22}.
Viral escape from neutralising antibodies can facilitate
breakthrough infections in vaccinated and convalescent individuals;
however, pre-existing cellular and innate immunity could protect from
severe disease \cite{Fleming22,Carreno-et-al22}.
Mutations in Omicron can knock out or substantially reduce neutralization
by most of the large panel of potent monoclonal antibodies and antibodies
under commercial development.
Studies also showed that neutralizing antibody titers against BA.2 were
similar to those against the BA.1 variant.
A third dose of the vaccine was needed for induction of consistent neutralizing
antibody titers against either the BA.1 or BA.2.3,4 variants,
suggesting a substantial degree of cross-reactive natural immunity
\cite{Yu-et-al22}.

The studies above indicate that vaccine immunity was lost substantially against Omicron, and natural immunity provided
better protection. 
All the studies were done in laboratories or in a small section of population, and our analysis in this section complements
them as it is based on population-wise data.

\subsection{Vaccine Immunity before Omicron}

We first analyze the gain in immunity due to vaccination before the arrival of Omicron. For this propose, we
downloaded an extensive list of serosurveys, carried out in various countries and
maintained by the site~\cite{serotracker},
eliminated surveys that were not done at national level, or had small
sample sizes, or had high risk of bias.
Nineteen countries remained after this pruning.
These sero-surveys were used together with
the SUTRA model to capture the pandemic trajectories and estimate parameter values in
these countries. We identify two values for each country:
\begin{enumerate}
\item Value $\frac{1}\r-1$ before arrival of Omicron. All the nineteen countries had restrictions removed well before Omicron and
therefore, it is reasonable to expect that $\ra$ was close to $1$ by the arrival of Omicron in the country. As shown in
Lemma~\ref{l-gain-loss}, $\r = \ra + \rloss - \rgain \approx 1 + \rloss - \rgain$ at the time. In other words,
$\rgain -\rloss \approx 1 - \r$.
Since calibration of the model provides only an approximate value of $\r$,
we use fractional gain in immunity $\frac{\rgain-\rloss}{\r} \approx \frac{1}{\r}-1$ which is likely to be more robust.
\item Fraction of uninfected population in the country that has received at least
one dose of vaccination at the onset of Omicron wave. To estimate this number,
we assume that the two types of immunity, vaccine and natural, are independent random variables,
implying that the fraction with hybrid immunity is
the product of vaccine immunity and natural immunity fractions.
With this assumption, and using vaccination data from~\cite{Ourworld}, we can estimate the required value.
\end{enumerate}

Figure~\ref{fig:Vaccine-Immunity} plots the above two numbers. It shows a very strong correlation between the two numbers implying that vaccination provided excellent immunity before Omicron.

\bfig
\bc
\includegraphics[width=160mm]{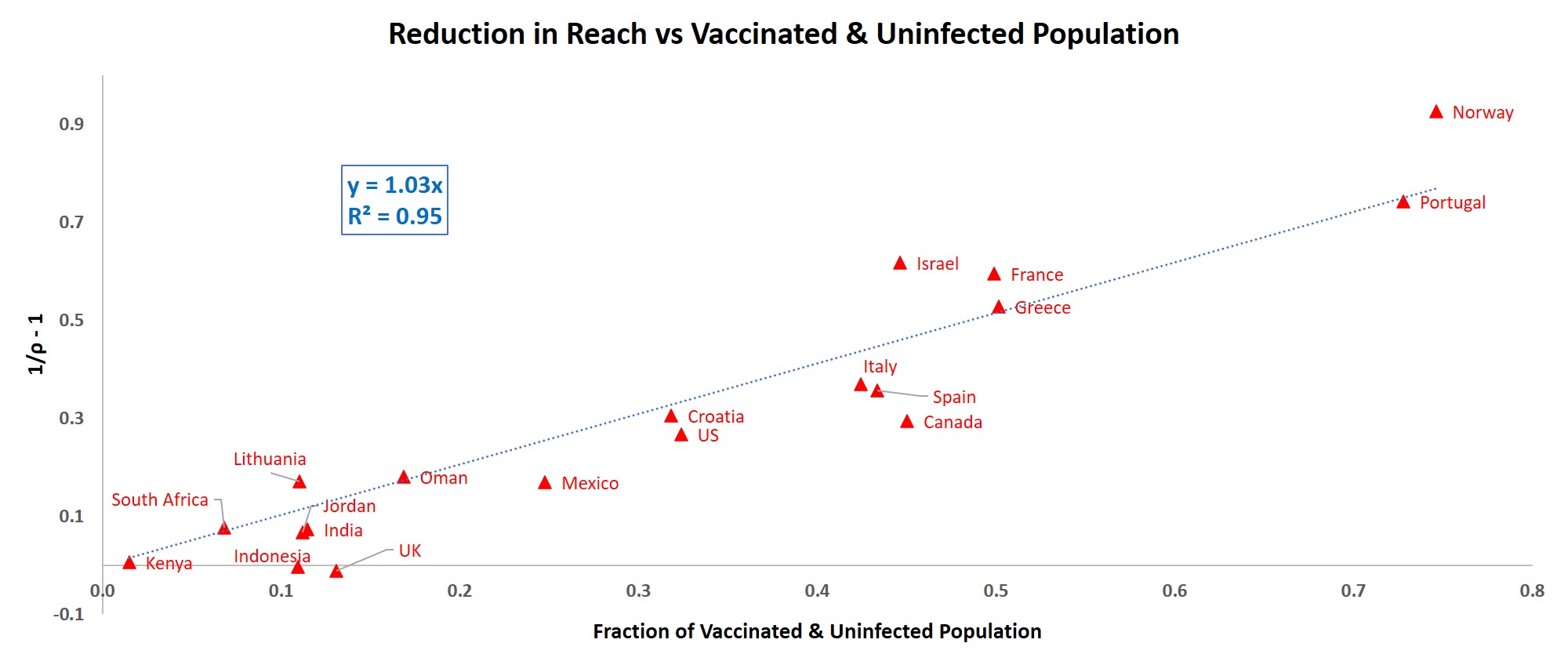}
\ec
\caption{$\frac{1}{\r}-1$ vs.\ Percentage of Vaccinated \&\ Uninfected Population}
\label{fig:Vaccine-Immunity}
\efig

\subsection{Loss of Vaccine Immunity after Omicron}

We can measure immunity loss due to Omicron by comparing the value of $\r$ after Omicron arrives in a country with the value 
before its arrival. This change will be almost entirely due to immunity loss since $\ra \approx 1$ before Omicron as discussed above. 

We first compare it with vaccine-only immunity present in the population to get an estimate of how much of it was lost.
Figure~\ref{fig:Immunity-Loss-after-Omicron} plots these two numbers. Again, a very strong correlation is observed between
the numbers. This, and the fact that the slope of best-fit line is close to $1$, suggests that almost all of immunity loss was
due to loss of vaccination immunity.

\bfig
\bc
\includegraphics[width=160mm]{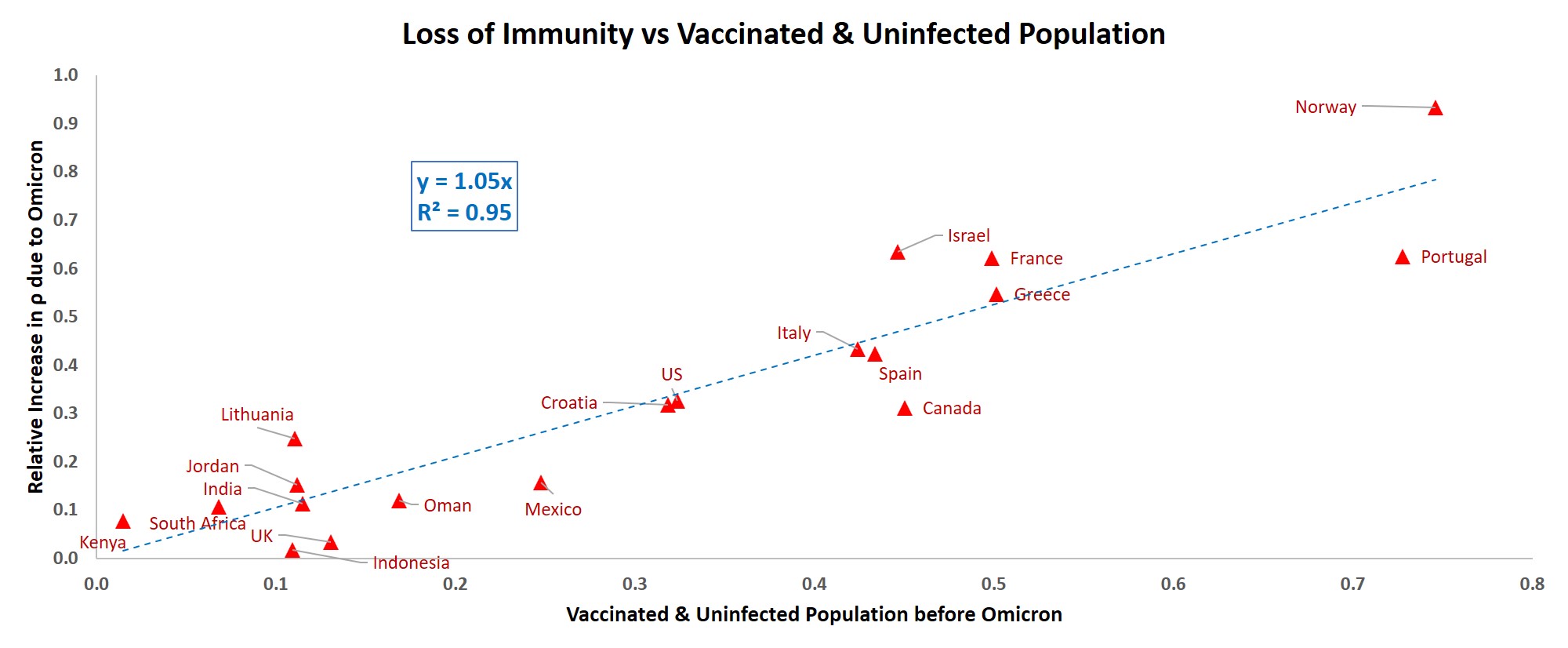}
\ec
\caption{Immunity Loss after Omicron vs.\ Vaccinated \&\ Uninfected Population}
\label{fig:Immunity-Loss-after-Omicron}
\efig

The above conclusion is further strengthened by the next plot where we compare immunity loss due to Omicron with the natural immunity present in the population before the arrival
of mutation. Figure~\ref{fig:Natural-Immunity-after-Omicron} plots these two numbers. It shows a very strong
negative correlation implying that natural immunity provided excellent protection against Omicron.

\bfig
\bc
\includegraphics[width=160mm]{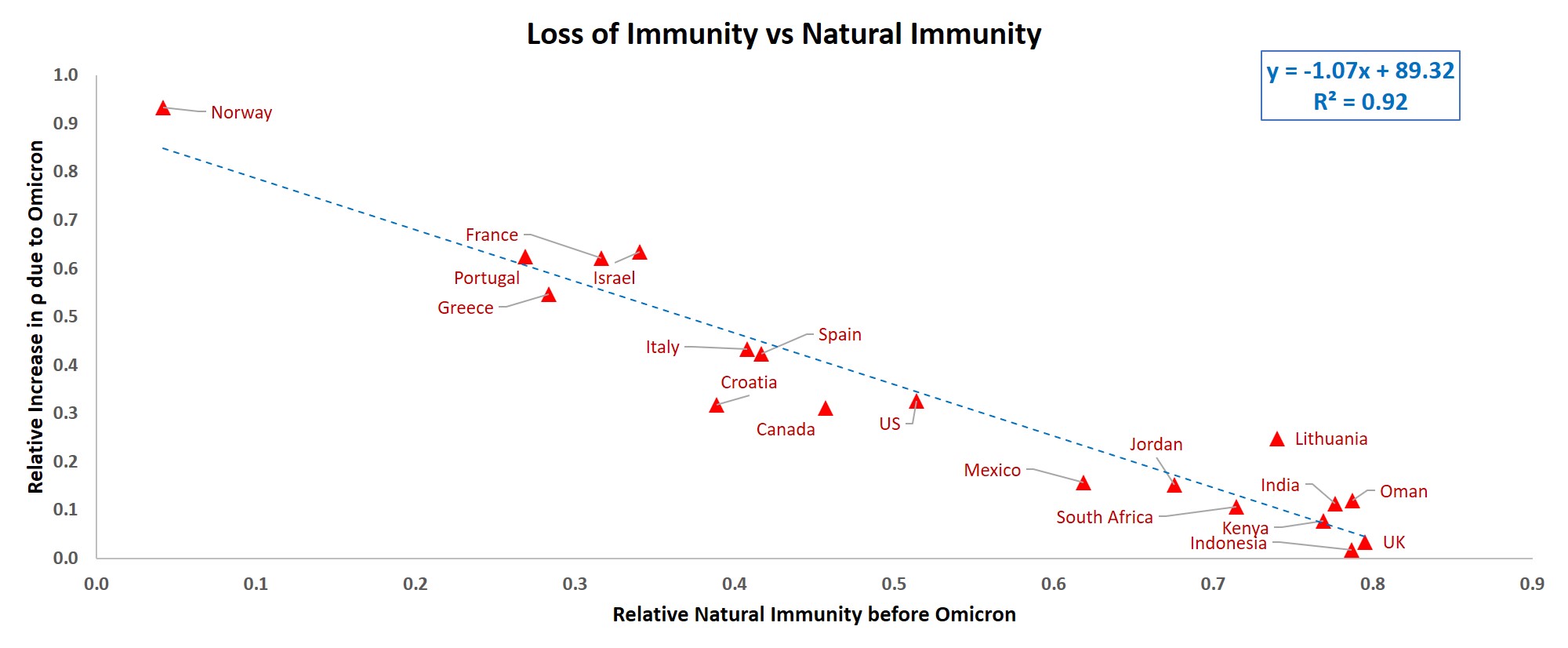}
\ec
\caption{Immunity Loss after Omicron vs.\ Naturally Immune Population}
\label{fig:Natural-Immunity-after-Omicron}
\efig

\subsection{Incorporating More Countries}

To make our conclusions more broad-based, we include seventeen more countries based on following criteria:
\begin{enumerate}
\item
All continents are represented well (five from Africa, two from North America, four from South America, thirteen from Asia, eleven from Europe, and one from Australia)
\item
Populous countries are simulated (except China for which it is not possible to calibrate the model). More than half the world's population lives in these countries.
\item
It is likely that $\ra$ was close to maximum in these countries at the time of Omicron's arrival, allowing us to calibrate the model.
\end{enumerate}

Adding these countries to the plots, we find little change in the correlations (see Figures~\ref{fig:Vaccine-Immunity-All},
\ref{fig:Immunity-Loss-after-Omicron-All}, \ref{fig:Natural-Immunity-after-Omicron-All}), further strengthening 
the conclusions.

\bfig
\bc
\includegraphics[width=160mm]{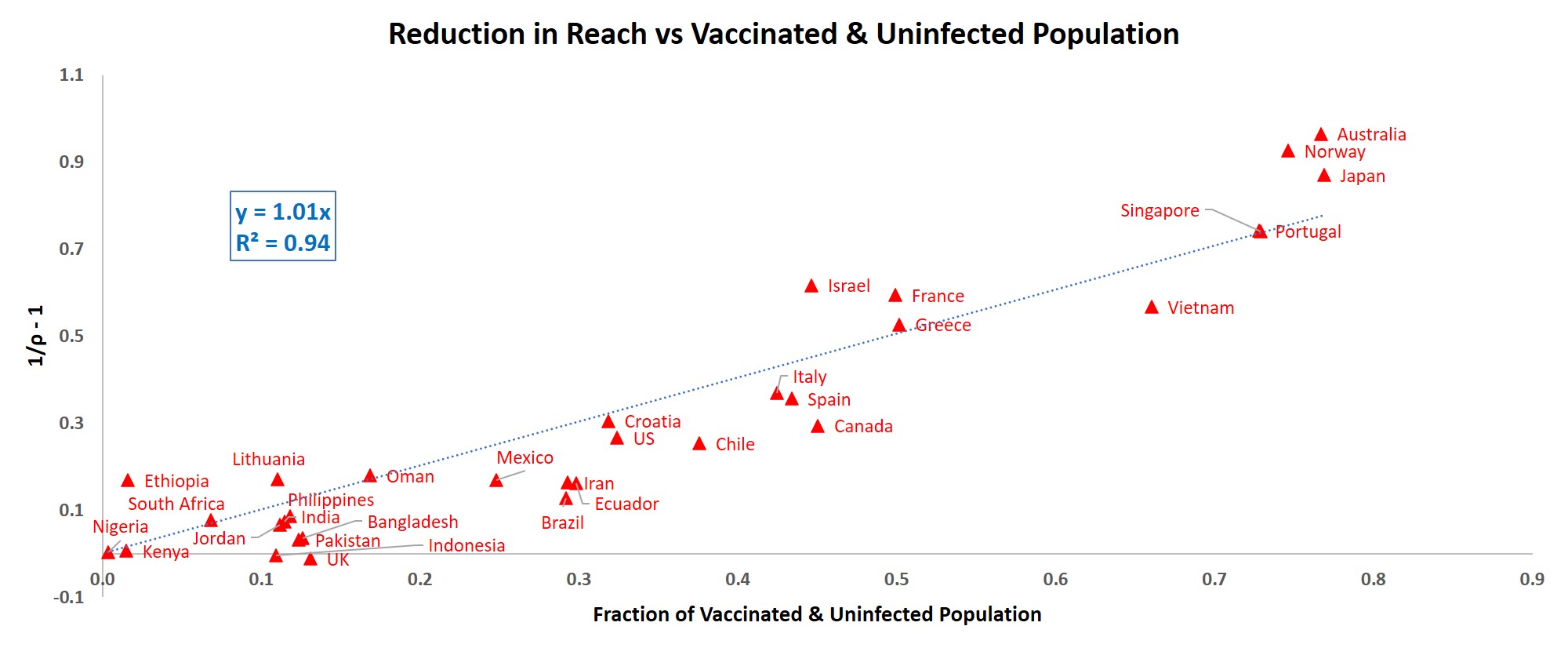}
\ec
\caption{$\frac{1}{\r}-1$ vs.\ Percentage of Vaccinated \&\ Uninfected Population}
\label{fig:Vaccine-Immunity-All}
\efig

\bfig
\bc
\includegraphics[width=160mm]{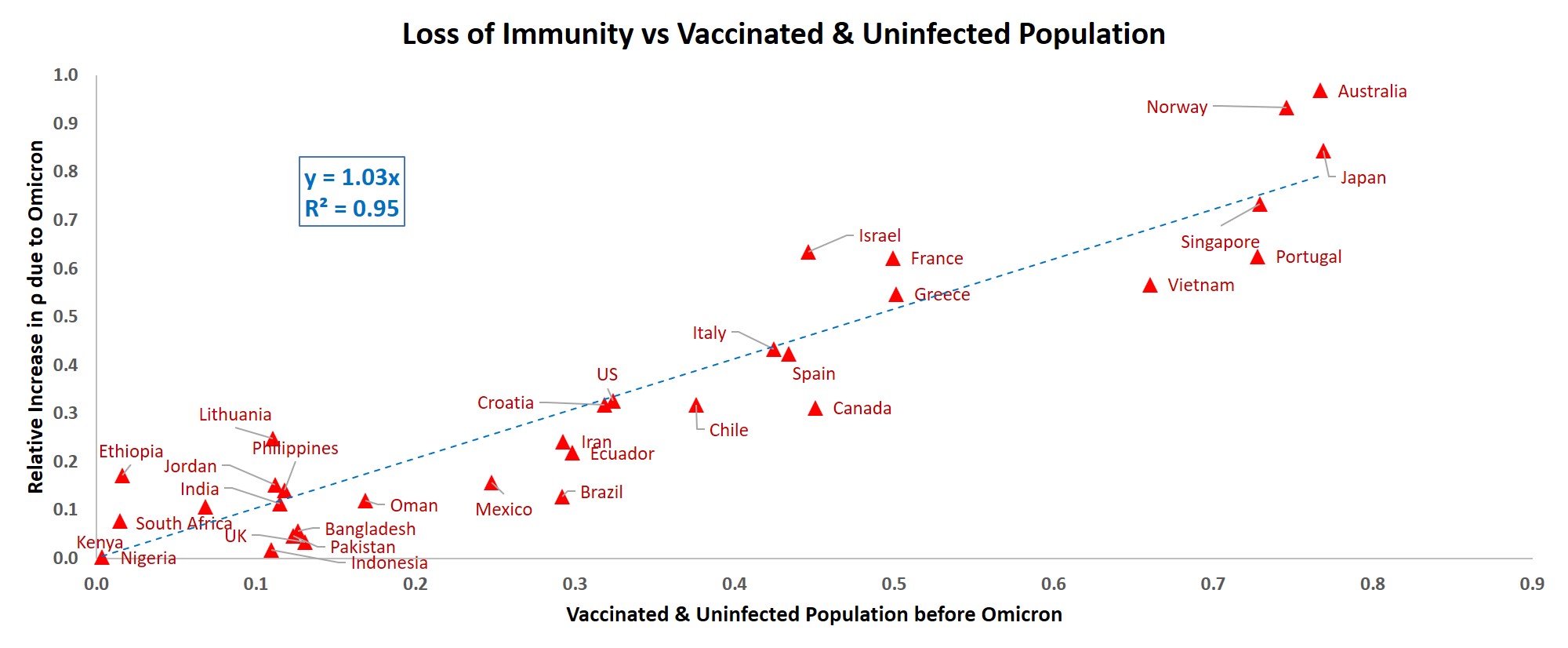}
\ec
\caption{Immunity Loss after Omicron vs.\ Vaccinated \&\ Uninfected Population}
\label{fig:Immunity-Loss-after-Omicron-All}
\efig

\bfig
\bc
\includegraphics[width=160mm]{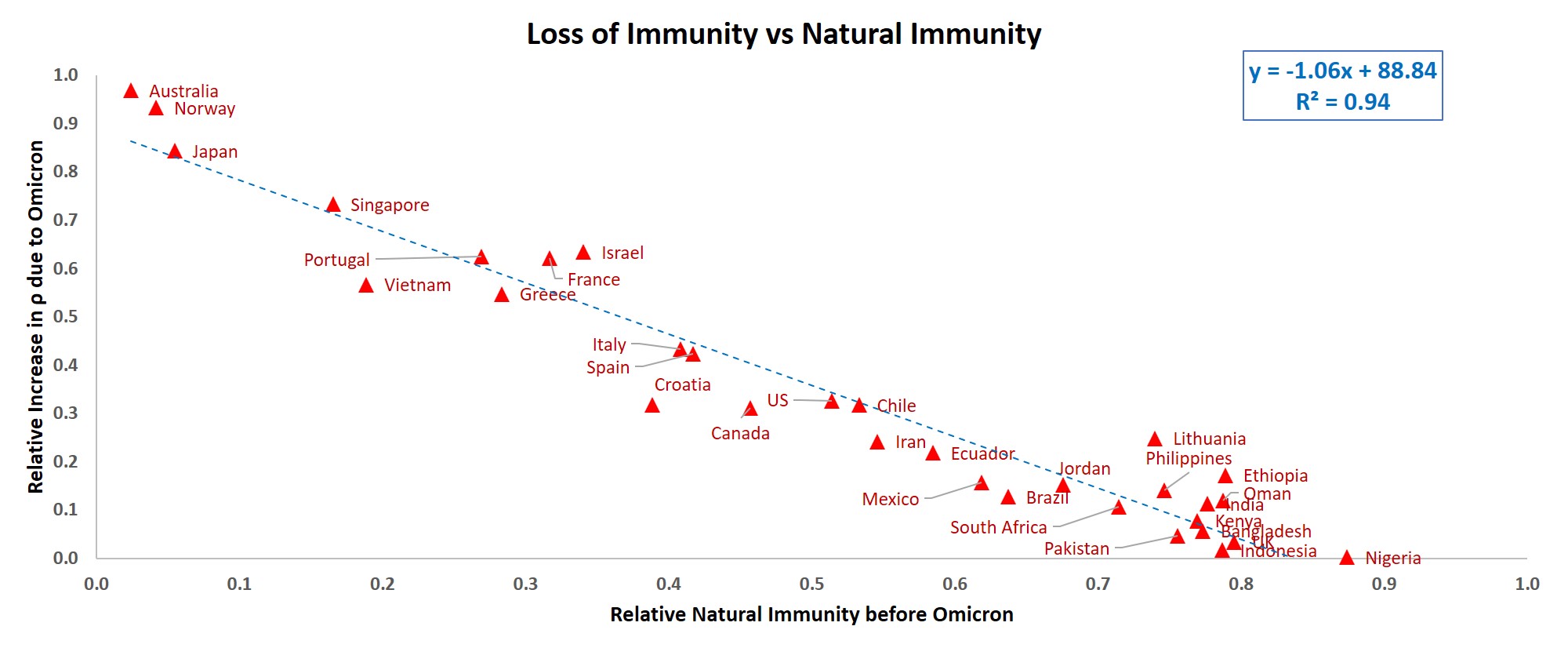}
\ec
\caption{Immunity Loss after Omicron vs.\ Naturally Immune Population}
\label{fig:Natural-Immunity-after-Omicron-All}
\efig

Taken together, these plots show that Omicron bypassed vaccination immunity almost completely, but natural immunity provided excellent protection. A clear conclusion is that countries that followed zero-COVID strategy -- strictly control the spread and vaccinate entire population -- suffered maximum during the Omicron wave. Indeed,
a perusal of Table \ref{table:Imm} shows that in countries
where the reach was very low before the arrival of the Omicron variant
saw very large percentage increases in reach thereafter.
It also suggests that the best strategy for managing the pandemic for a country is to allow the virus to freely spread after vaccinating entire population. Trying to control its spread even after vaccination will not build natural immunity in the population
and there will always be a chance of fresh outbreaks. We can see it happening in China at present.

\clearpage

\section*{Acknowledgments}

The work of MA, DP, TH, Arti S, Avaneesh S, \&\ Prabal S was supported by grants from CII and Infosys Foundation, and MV was supported by the Science and Engineering Research
Board, India.


\begin{thebibliography}{99}

\bibitem{worldometers}
Worldometers. 2022  {COVID-19 Coronavirus Pandemic}.
  https://www.worldometers.info/coronavirus/.

\bibitem{Honigsbaum20}
Honigsbaum M. 2020  Revisiting the 1957 and 1968 influenza pandemics. {\em The
  Lancet} \textbf{395}, 1824--1826.

\bibitem{CDC-Spanish}
for Disease~Control C, Prevention. 2019  1918 Pandemic (H1N1 virus).
  https://www.cdc.gov/flu/pandemic-resources/1918-pandemic-h1n1.html.

\bibitem{math-models}
{Siettos CI, Russo L}. 2013  {Mathematical modeling of infectious disease
  dynamics}. {\em Virulence} \textbf{4}, 295--306.

\bibitem{Ker-McK27}
Kermack WO, McKendrick AG. 1927  A contribution to the mathematical theory of
  epidemics. {\em Proceedings of The Royal Society {A}} \textbf{117}, 700--721.

\bibitem{Hethcote76}
Hethcote HW. 1976  Qualitative analyses of communicable disease models. {\em
  Mathematical Biosciences} \textbf{28}, 335--356.

\bibitem{Rob-Sti13}
Robinson M, Stilianakis NI. 2013  A model for the emergence of drug resistance
  in the presence of asymptomatic infections. {\em Mathematical Biosciences}
  \textbf{243}, 163--177.

\bibitem{SIDARTHE}
{Giordano, G., Blanchini, F., Bruno, R. et al.}. 2020  {Modelling the COVID-19
  epidemic and implementation of population-wide interventions in Italy}. {\em
  {Nat Med}} \textbf{26}, 855--860.

\bibitem{SAIR-AOC}
MartÃ­nez-Guerra R, Flores-Flores JP. 2021  An algorithm for the robust
  estimation of the COVID-19 pandemicâ€™s population by considering undetected
  individuals. {\em Applied Mathematics and Computation} \textbf{405}, 126273.

\bibitem{SUTR-posterior}
Deo V, Grover G. 2021  A new extension of state-space SIR model to account for
  Underreporting â€“ An application to the COVID-19 transmission in California
  and Florida. {\em Results in Physics} \textbf{24}, 104182.

\bibitem{SIRD-time-varying}
{Calafiore G C and Novara C and Possieri C}. 2020  {A time-varying SIRD model
  for the COVID-19 contagion in Italy}. {\em {Annual Rev Control}} \textbf{50},
  361--372.

\bibitem{MV-IJMR20}
Agrawal M, Kanitkar M, Vidyasagar M. 2020  Modelling the spread of {SARS-CoV-2}
  pandemic - Impact of lockdowns \& interventions. {\em Indian Journal of
  Medical Research}.

\bibitem{infection-duration}
{Byrne AW, McEvoy D, Collins AB, et al}. 2020  {Inferred duration of infectious
  period of SARS-CoV-2: rapid scoping review and analysis of available evidence
  for asymptomatic and symptomatic COVID-19 cases}. {\em BMJ Open} \textbf{10}.

\bibitem{infection-duration-2}
{Daniel Owusu, Mary A Pomeroy, Nathaniel M Lewis et al}. 2021  {Persistent
  SARS-CoV-2 RNA Shedding Without Evidence of Infectiousness: A Cohort Study of
  Individuals With COVID-19}. {\em The Journal of Infectious Diseases}.

\bibitem{India-tweet-delta}
{Manindra Agrawal}. 2021  {Tweet on 29th April}.
  \url{https://twitter.com/agrawalmanindra/status/1387734516807073792?s=20}.

\bibitem{India-tweet-omicron}
{Manindra Agrawal}. 2022  {Tweet on 13th February}.
  \url{https://twitter.com/agrawalmanindra/status/1492569073783570432?s=20\&t=ULH1TMDIArosjgjZLvvRXw}.

\bibitem{UK-tweet-delta}
{Manindra Agrawal}. 2021a  {Tweet on 10th July}.
  \url{https://twitter.com/agrawalmanindra/status/1413762687213789185?s=20\&t=ULH1TMDIArosjgjZLvvRXw}.

\bibitem{US-tweet-delta}
{Manindra Agrawal}. 2021b  {Tweet on 25th July}.
  \url{https://twitter.com/agrawalmanindra/status/1419176814954500096?s=20}.

\bibitem{covid19India}
{COVID-19 India}. 2021  {COVID-19 India}. https://www.covid19india.org/.

\bibitem{India-serosurvey-three}
{Manoj V Murhekar}, {Jeromie Wesley Vivian Thangaraj} T et~al.. 2021a
  {SARS-CoV-2 seroprevalence among the general population and healthcare
  workers in India}. {\em {Int J Infect Dis.}}

\bibitem{India-serosurvey-two}
{Manoj V Murhekar}, {Tarun Bhatnagar}, {Sriram Selvaraju}, {V Saravanakumar},
  {Jeromie Wesley Vivian Thangaraj} et~al.. 2021b  {SARS-CoV-2 antibody
  seroprevalence in India, Augustâ€“September, 2020: findings from the second
  nationwide household serosurvey}. {\em {LANCET Global Health}} \textbf{9},
  E257--E266.

\bibitem{India-serosurvey-four}
{Manoj V Murhekar}, {Tarun Bhatnagar}, {Jeromie Wesley Vivian Thangaraj}
  et~al.. 2021c  {Seroprevalence of IgG antibodies against SARS-CoV-2 among the
  general population and healthcare workers in India, Juneâ€“July 2021: A
  population-based cross-sectional study}. {\em {PLOS Medicine}}.

\bibitem{timeline-India}
{Wikipedia}. 2022  {Timeline of the COVID-19 pandemic in India}.
  \url{https://en.wikipedia.org/wiki/Timeline_of_the_COVID-19_pandemic_in_India}.

\bibitem{UK-serosurvey}
Ward H, Cooke G, Atchison C, Whitaker M et~al.. 2021  {Prevalence of antibody
  positivity to SARS-CoV-2 following the first peak of infection in England:
  Serial cross-sectional studies of 365,000 adults}. {\em {LANCET Regional
  Health}}.

\bibitem{timeline-UK}
{Wikipedia}. 2022  {Timeline of the COVID-19 pandemic in the United Kingdom}.
  \url{https://en.wikipedia.org/wiki/Timeline_of_the_COVID-19_pandemic_in_the_United_Kingdom}.

\bibitem{US-serosurvey}
{Robert Stout and Steven Rigatti}. 2021  {Seroprevalence of SARS-CoV-2
  Antibodies in the US Adult Asymptomatic Population as of September 30, 2020}.
  {\em {JAMA Newwork Open}}.

\bibitem{timeline-US}
{Wikipedia}. 2022  {Timeline of the COVID-19 pandemic in the United States}.
  \url{https://en.wikipedia.org/wiki/Timeline_of_the_COVID-19_pandemic_in_the_United_States}.

\bibitem{SA-serosurvey}
{Nicole Wolter}, {Stefano Tempia}, {Anne von Gottberg}, {Jinal Bhiman} et~al..
  2022  {Seroprevalence of SARS-CoV-2 after the second wave in South Africa in
  HIV-infected and uninfected persons: a cross-sectional household survey}.
  {\em {Clinical Infectious Diseases}}.

\bibitem{timeline-SA}
{Wikipedia}. 2022  {Timeline of the COVID-19 pandemic in South Africa}.
  \url{https://en.wikipedia.org/wiki/Timeline_of_the_COVID-19_pandemic_in_South_Africa}.

\bibitem{Denji-et-al22}
Dejnirattisai W, Huo J, Zhou D et~al.. 2022  {SARS-CoV-2 Omicron-B.1.1.529
  leads to widespread escape from neutralizing antibody responses}. {\em Cell}
  \textbf{185}, 467--484.e15.

\bibitem{Zhang-et-al22}
Zhang L, Li Q, Liang Z et~al.. 2022  {The significant immune escape of
  pseudotyped SARS-CoV-2 variant Omicron}. {\em Emerging Microbes \&
  Infections} \textbf{11}, 1--5.

\bibitem{Ren-et-al22}
Ren SY, Wang WB, Gao RD, Zhou AM. 2022  {Omicron variant (B.1.1.529) of
  SARS-CoV-2: Mutation, infectivity, transmission, and vaccine resistance}.
  {\em World Journal of Clinical Cases} \textbf{10}, 1--11.

\bibitem{Cele-et-al21}
Cele S, Jackson L, Khoury DS et~al.. 2021  {SARS-CoV-2 Omicron has extensive
  but incomplete escape of Pfizer BNT162b2 elicited neutralization and requires
  ACE2 for infection}. {\em Nature}.

\bibitem{Zeng-et-al21}
Zeng C, Evans JP, Qu P et~al.. 2021  {Neutralization and Stability of
  SARS-CoV-2 Omicron Variant}. {\em bioRxiv}.

\bibitem{Ai-et-al22}
Ai J, Zhang H, Zhang Y et~al.. 2022  {Omicron variant showed lower neutralizing
  sensitivity than other SARS-CoV-2 variants to immune sera elicited by
  vaccines after boost}. {\em Emerging Microbes \& Infections} \textbf{11},
  337--343.

\bibitem{Garcia-et-al22}
{St. Denis} WFGBKJ, Hoelzemer A et~al.. 2022  {mRNA-based COVID-19 vaccine
  boosters induce neutralizing immunity against SARS-CoV-2 Omicron variant}.
  {\em Cell} \textbf{185}, 457--466.e4.

\bibitem{Planas-et-al22}
Planas D, Saunders N, Maes P et~al.. 2022  {Considerable escape of SARS-CoV-2
  Omicron to antibody neutralization}. {\em Nature} \textbf{602}, 671--675.

\bibitem{Fleming22}
Flemming A. 2022  Omicron, the great escape artist. {\em Nature Reviews
  Immunology} \textbf{22}, 75.

\bibitem{Carreno-et-al22}
{n}o JMC, Alshammary H, Tcheou J et~al.. 2022  {Activity of convalescent and
  vaccine serum against SARS-CoV-2 Omicron}. {\em Nature} \textbf{602},
  682--688.

\bibitem{Yu-et-al22}
Yu J, Collier AY, Rowe M et~al.. 2022  {Neutralization of the SARS-CoV-2
  Omicron BA.1 and BA.2 Variants}. {\em New England Journal of Medicine}
  \textbf{386}, 1579--1580.

\bibitem{serotracker}
Arora RK, Joseph A, Wyk JV et~al.. 2022  Serotracker.
  https://serotracker.com/en.

\bibitem{Ourworld}
{Our World in Data}. 2022  Vaccination Statistics.
  https://ourworldindata.org/covid-vaccinations.

\end{thebibliography}

\newpage 

\appendix

\section{India Phase 9 Plots} \label{sec-phase-plots}

All the plots in this section have $\frac{1}{P_0} \aC_T \aT$ on $x$-axis and $\aT - \frac{1}{\bt} \aN_T$ on $y$-axis.

\bfig[h!]
\bc
\begin{minipage}[b]{0.45\linewidth}
\includegraphics[width=75mm]{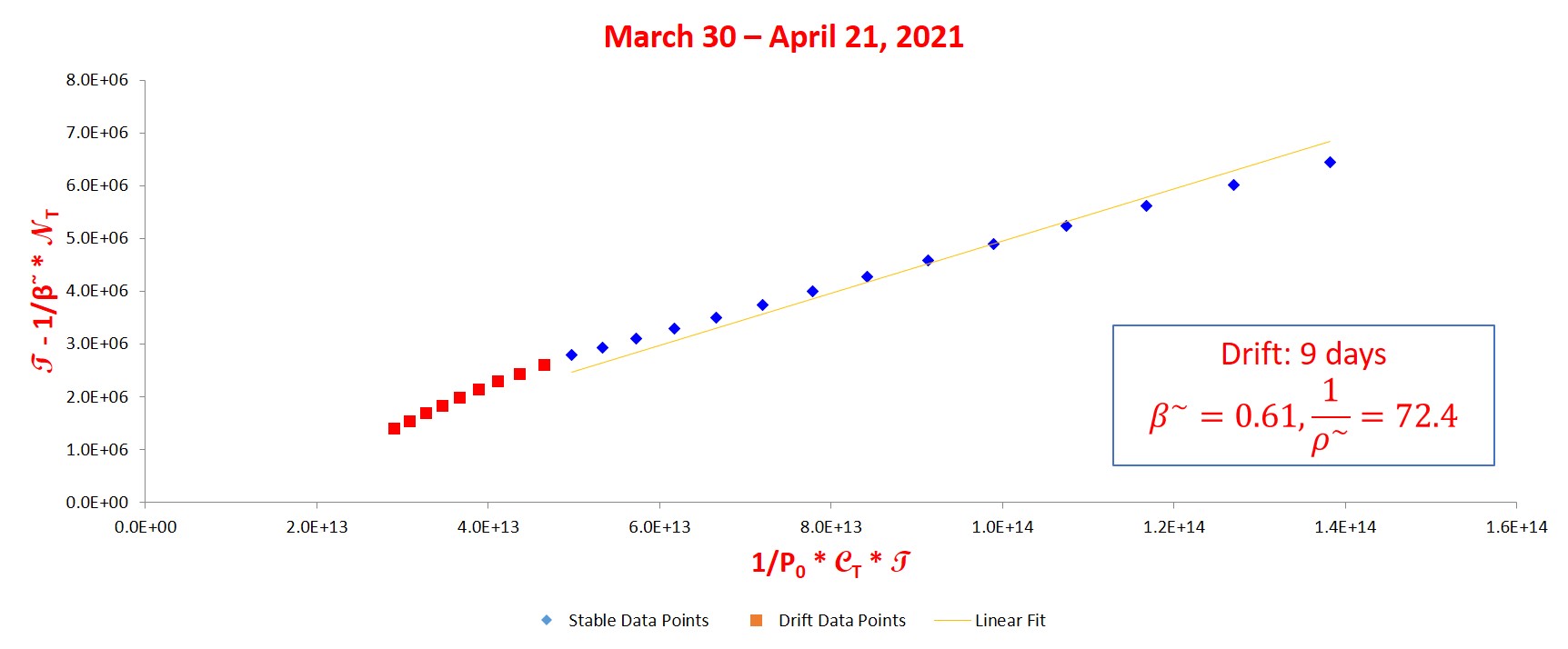}
\caption{Phase Plot on April 21, 2021}
\label{fig:IN-Ph-04-21}
\end{minipage}
\quad
\begin{minipage}[b]{0.45\linewidth}
\includegraphics[width=75mm]{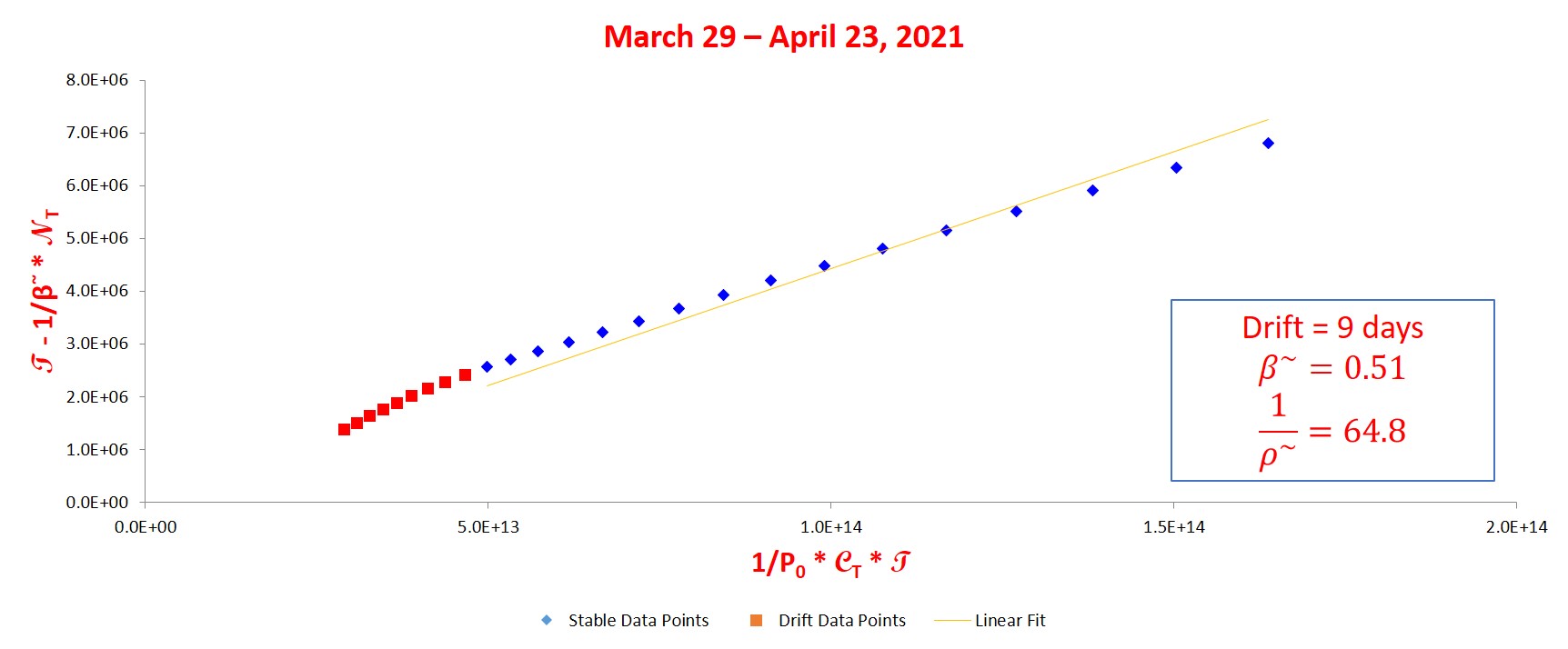}
\caption{Phase Plot on April 23, 2021}
\label{fig:IN-Ph-04-23}
\end{minipage}
\ec
\efig

\bfig[h!]
\bc
\begin{minipage}[b]{0.45\linewidth}
\includegraphics[width=75mm]{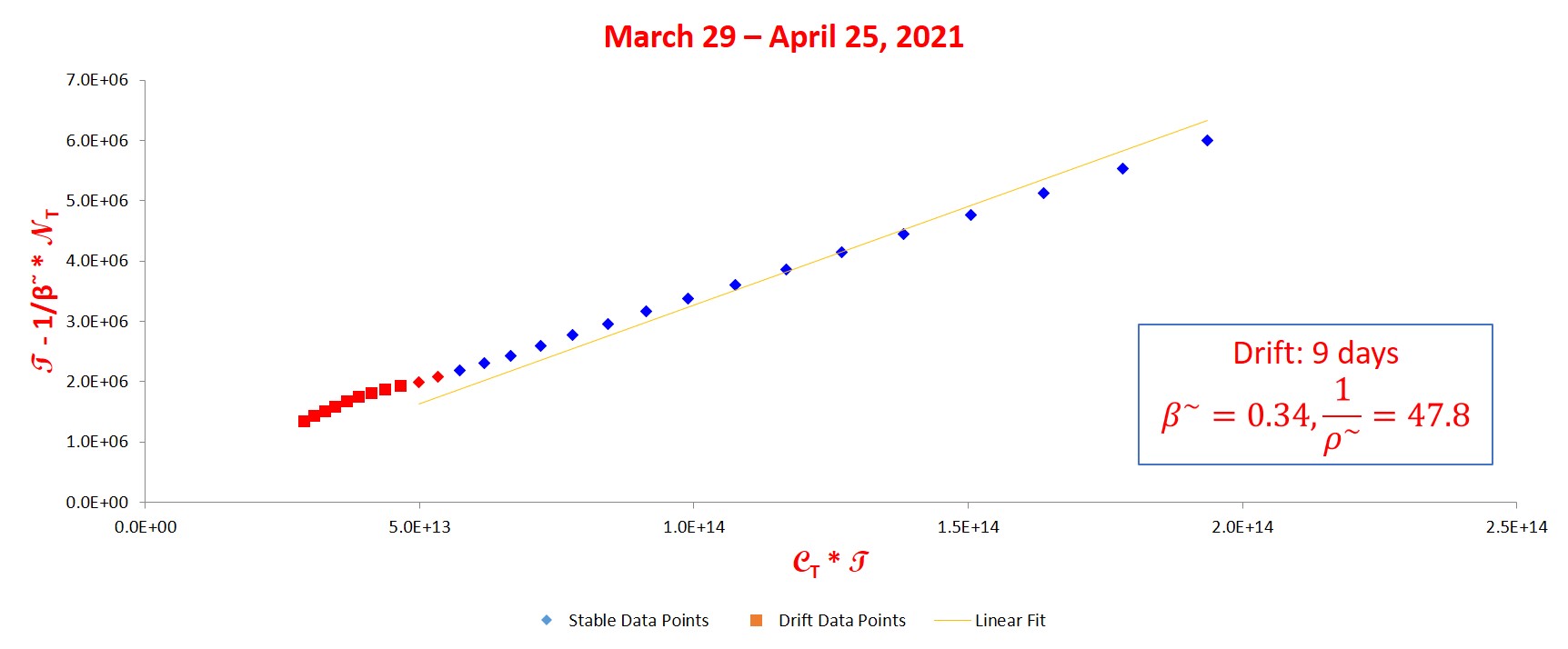}
\caption{Phase Plot on April 25, 2021}
\label{fig:IN-Ph-04-25}
\end{minipage}
\quad
\begin{minipage}[b]{0.45\linewidth}
\includegraphics[width=75mm]{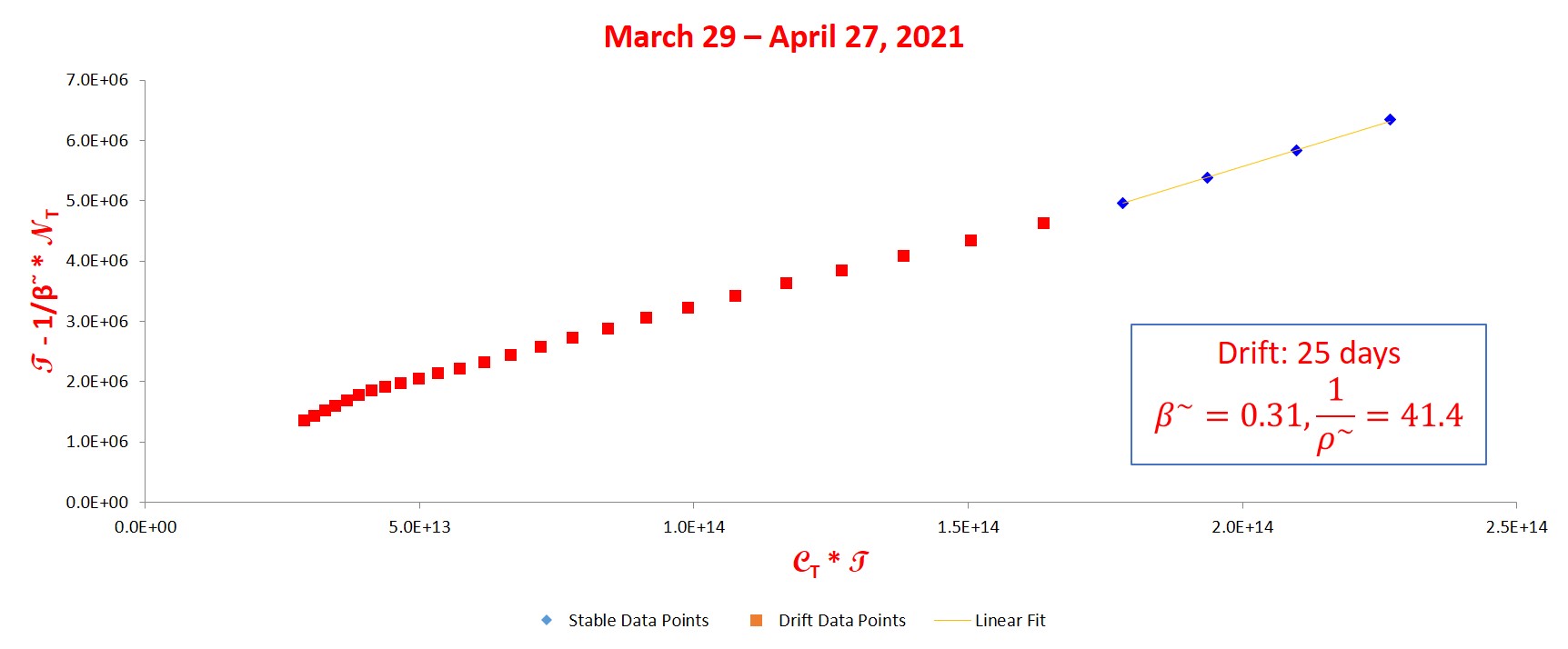}
\caption{Phase Plot on April 27, 2021}
\label{fig:IN-Ph-04-27}
\end{minipage}
\ec
\efig

\bfig[h!]
\bc
\begin{minipage}[b]{0.45\linewidth}
\includegraphics[width=75mm]{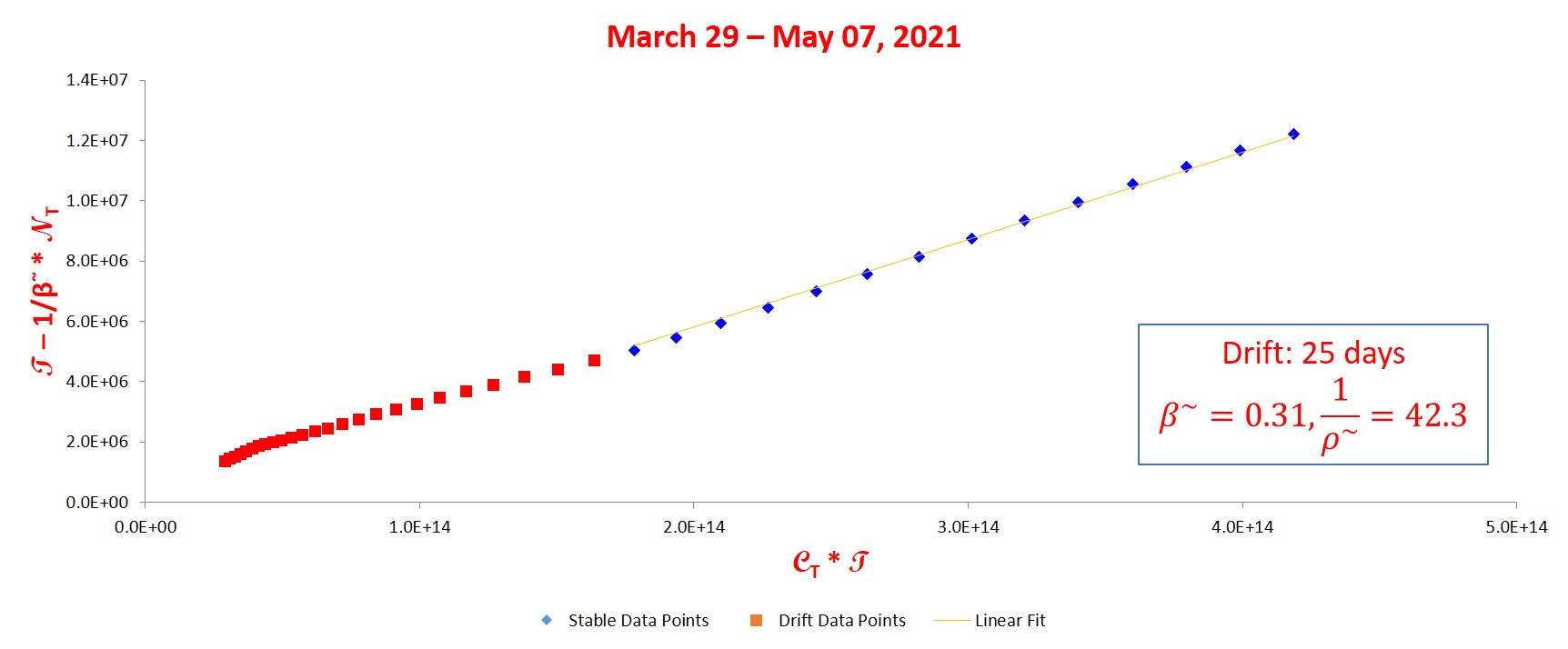}
\caption{Phase Plot on May 07, 2021}
\label{fig:IN-Ph-05-07}
\end{minipage}
\quad
\begin{minipage}[b]{0.45\linewidth}
\includegraphics[width=75mm]{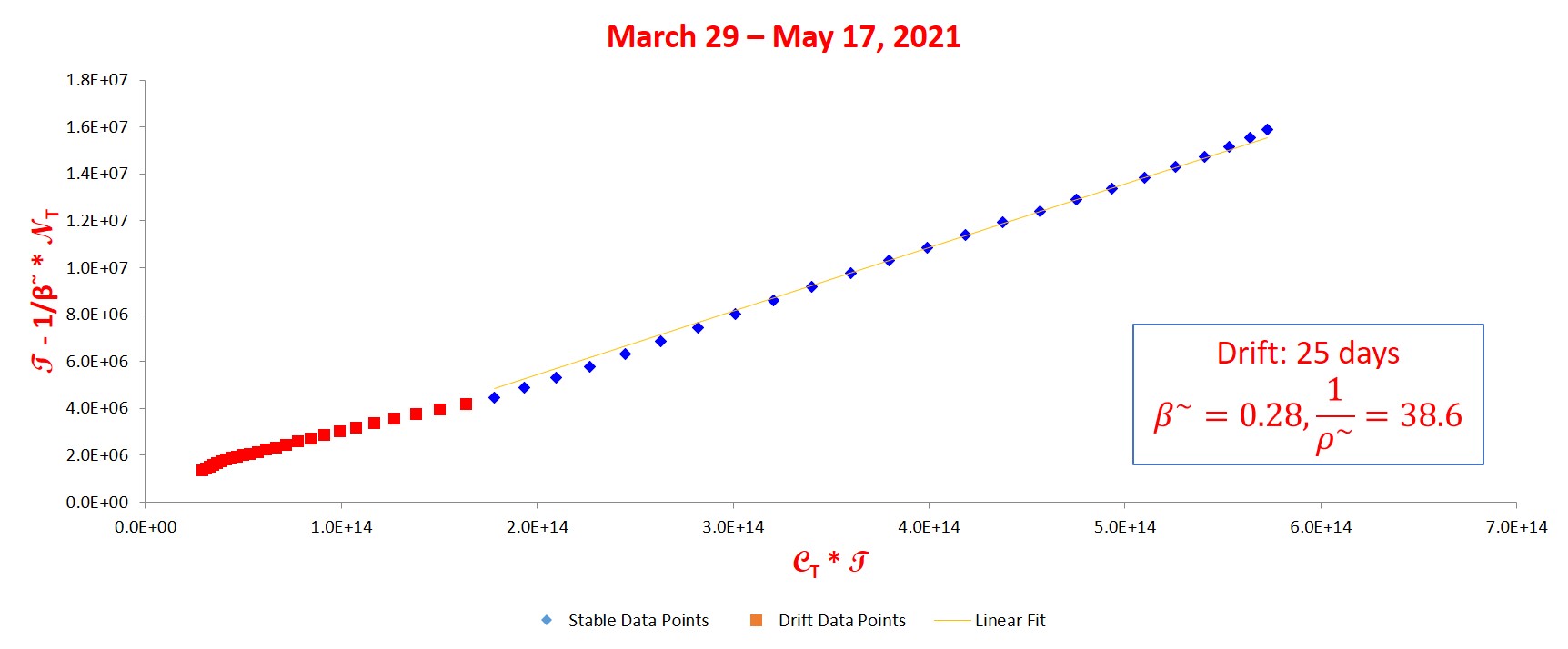}
\caption{Phase Plot on May 17, 2021}
\label{fig:IN-Ph-05-17}
\end{minipage}
\ec
\efig

\bfig[h!]
\bc
\begin{minipage}[b]{0.45\linewidth}
\includegraphics[width=75mm]{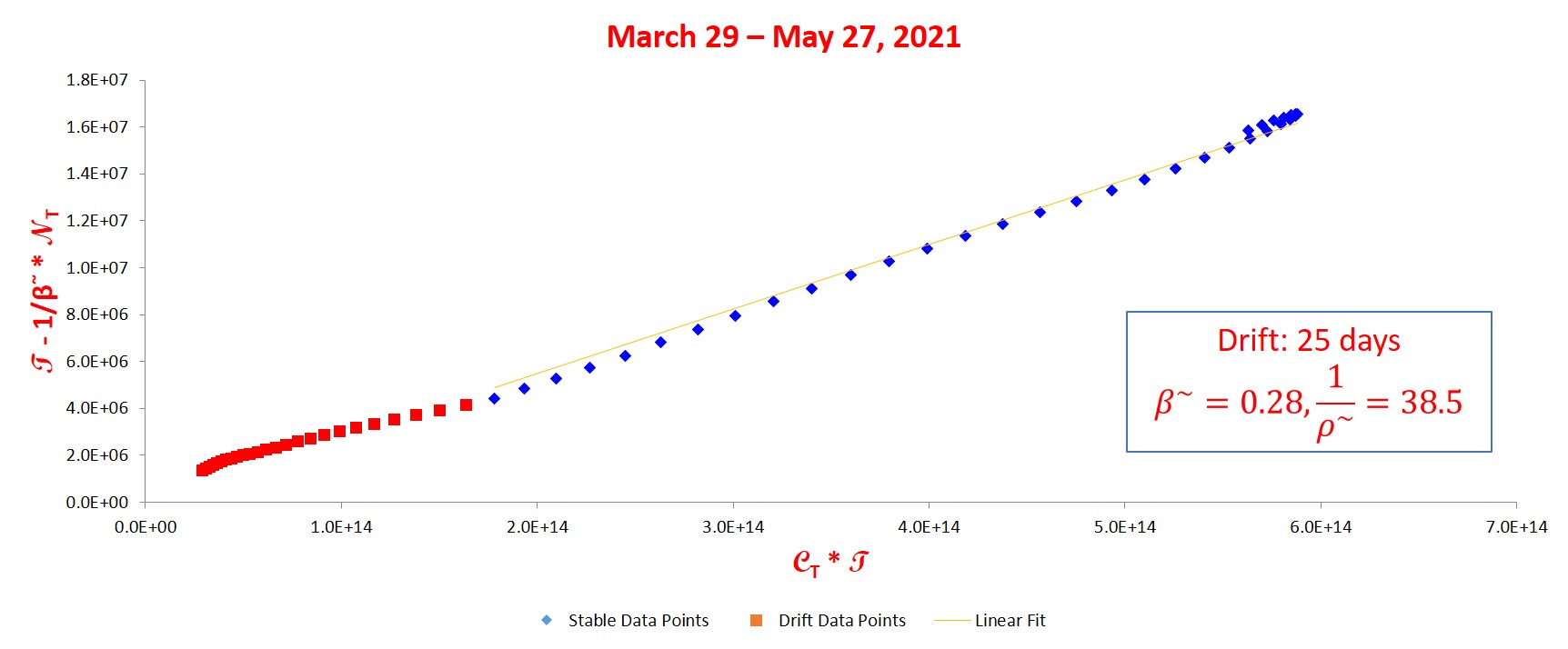}
\caption{Phase Plot on May 27, 2021}
\label{fig:IN-Ph-05-27}
\end{minipage}
\quad
\begin{minipage}[b]{0.45\linewidth}
\includegraphics[width=75mm]{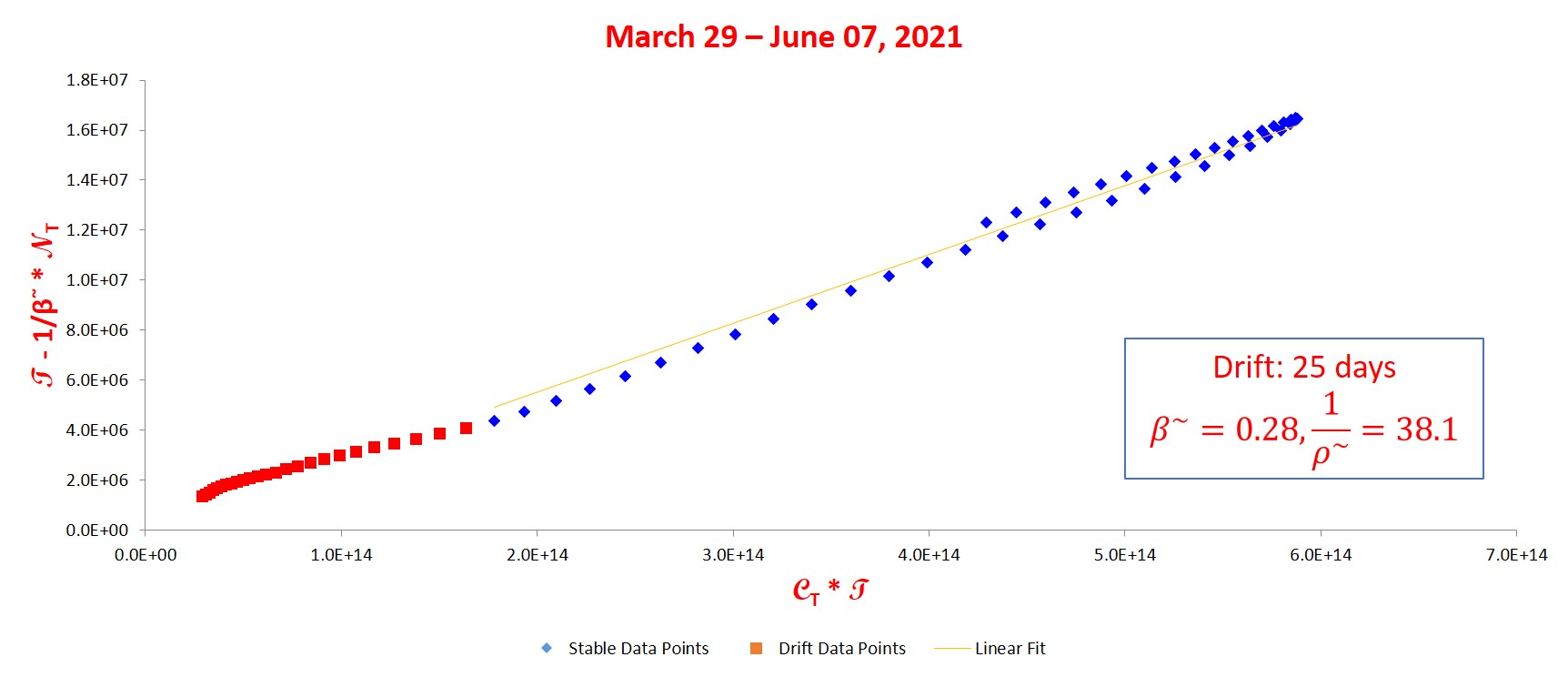}
\caption{Phase Plot on June 07, 2021}
\label{fig:IN-Ph-06-07}
\end{minipage}
\ec
\efig

\clearpage

\section{Pandemic Status in Countries at the Time of Omicron Arrival}\label{sec-tables}

In this section, we provide the data used for analysis in section~\ref{sec-immunity}. The table below lists, for $32$ countries, the percentage of population that had vaccine immunity (given at least one shot of vaccine fifteen days before) when Omicron arrived in the country. It also lists percentage of population with natural immunity (estimated by the model), with hybrid immunity (assuming two types of immunity are independent), with vaccine-only immunity (difference vaccine immunity and hybrid immunity), and within reach of the pandemic at the time.

\begin{table}[h!]
\caption{Status of Pandemic at the Time of Omicron Arrival}
\bc
\btab{|l|c|r|r|r|r|r|}
\hline
& Serosurvey &  Vaccination & Natural & Hybrid & Only & Increase \\
Country & Available  & Immunity \% & Immunity \% & Immunity \% & Vaccination & in Reach \% \\
& & & & & Immunity \% & \\
\hline\hline
Australia	 & N	& 78.8	& 2.7	& 2.1	& 76.7	& 96.9 \\
\hline
Bangladesh & N	& 60.0	& 79.0	& 47.4	& 12.6	& 5.7 \\
\hline
Brazil	& N	& 80.4	& 63.7	& 51.2	& 29.2	& 12.8 \\
\hline
Canada	& Y	& 84.8	& 46.9	& 39.8	& 45.0	& 31.2 \\
\hline
Chile	& N	& 90.1	& 58.3	& 52.5	& 37.6	& 31.8 \\
\hline
Croatia	& Y	& 52.7	& 39.6	& 20.9	& 31.8	& 31.8 \\
\hline
Ecuador	& N	& 79.9	& 62.7	& 50.1	& 29.8	& 21.8 \\
\hline
Ethiopia	& N	& 7.7	& 79.2	& 6.1	& 1.6	& 17.2 \\
\hline
France	& Y	& 75.6	& 34.0	& 25.7	& 49.9	& 62.2 \\
\hline
Greece	& Y	& 71.2	& 29.6	& 21.1	& 50.1	& 54.7 \\
\hline
India	& Y	& 60.7	& 81.1	& 49.2	& 11.5	& 11.3 \\
\hline
Indonesia &Y & 55.5	& 80.3	& 44.6	& 10.9	& 1.7 \\
\hline
Iran	& N	& 73.5	& 60.2	& 44.2	& 29.3	& 24.2 \\
\hline
Israel	& Y	& 69.3	& 35.6	& 24.7	& 44.6	& 63.4 \\
\hline
Italy	 & Y 	& 77.7	& 45.4	& 35.3	& 42.4	& 43.3 \\
\hline
Japan & N	& 80.5	& 4.5	& 3.6	& 76.9	& 84.3 \\
\hline
Jordan	& Y	& 43.5	& 74.3	& 32.3	& 11.2	& 15.2 \\
\hline
Kenya	& Y	& 8.7	& 82.9	& 7.2	& 1.5	& 7.8 \\
\hline
Lithuania & Y & 60.0	& 81.6	& 49.0	& 11.0	& 24.8 \\
\hline
Mexico	& Y	& 63.5	& 61.0	& 38.7	& 24.8	& 15.7 \\
\hline
Nigeria	& N	& 2.8	& 87.3	& 2.4	& 0.4	& 0.3 \\
\hline
Norway	& Y	& 78.2	& 4.6	& 3.6	& 74.6	& 93.3 \\
\hline
Oman	& Y	& 63.1	& 73.3	& 46.3	& 16.8	& 11.9 \\
\hline
Pakistan &	N	& 53.0	& 76.7	& 40.7	& 12.3	& 4.6 \\
\hline
Philippines	& N	& 57.2	& 79.4	& 45.4	& 11.8	& 14.1 \\
\hline
Portugal	& Y	& 89.3	& 18.5	& 16.5	& 72.8	& 62.5 \\
\hline
Singapore	& N	& 86.9	& 16.1	& 14.0	& 72.9	& 73.3 \\
\hline
South Africa	& Y	& 26.0	& 73.8	& 19.2	& 6.8	& 10.7 \\
\hline
Spain	& Y & 81.2	& 46.6	& 37.8	& 43.4	& 42.4 \\
\hline
UK	& Y	& 78.3	& 83.3	& 65.2	& 13.1	& 3.4 \\
\hline
US	& Y	& 73.4	& 55.9	& 41.0	& 32.4	& 32.5 \\
\hline
Vietnam	& N	& 81.3	& 18.8	& 15.3	& 66.0	& 56.7 \\
\hline
\etab
\ec
\label{table:Imm}
\end{table}

\clearpage

\section{Proofs}\label{sec-proofs}

In this Appendix, we provide proofs of all lemmas and theorems stated in the paper.

\begin{oldLemma}{l-alternate-R}
When $\pvec{u}$ is independent of $\pvec{v}$ as well as $\pvec{w}$, there is a maxima of $R^2$ with $R^2_{\beta},
R^2_{\r}, \bt, \rt > 0$.
\end{oldLemma}

\begin{proof}
Let $x = \frac{1}{\bt}$ and $y = \frac{1}{\rt}$. Then we have:
\begin{equation}\label{eq:50}
\begin{split}
R^2 & = \frac{xy(2\pvec{v}^T\pvec{u}-y\pvec{v}^T\pvec{w}-x\pvec{v}^T\pvec{v})
(2\pvec{w}^T\pvec{u}-x\pvec{w}^T\pvec{v}-y\pvec{w}^T\pvec{w})}%
{|\pvec{u}-y\pvec{w}|^2 |\pvec{u}-x\pvec{v}|^2}
\end{split}
\end{equation}
with $R^2_{\beta} > 0$ iff $2\pvec{w}^T\pvec{u}-x\pvec{w}^T\pvec{v}-y\pvec{w}^T\pvec{w} > 0$
and $R^2_{\e} > 0$ iff $2\pvec{v}^T\pvec{u}-y\pvec{v}^T\pvec{w}-x\pvec{v}^T\pvec{v} > 0$.

The denominator of equation~(\ref{eq:50}) is always positive since
$\pvec{u}$ is independent of $\pvec{v}$ as well as $\pvec{w}$.
The numerator is a product of four linear terms in the
unknowns $x$ and $y$.
Therefore the value of $R^2$ is positive inside the
polygon defined by:
\begin{eqnarray*}
x & \geq & 0 \\
y & \geq & 0 \\
2\pvec{v}^T\pvec{u}-y\pvec{v}^T\pvec{w}-x\pvec{v}^T\pvec{v} & \geq & 0 \\
2\pvec{w}^T\pvec{u}-x\pvec{w}^T\pvec{v}-y\pvec{w}^T\pvec{w} & \geq & 0
\end{eqnarray*}
and is zero on the boundaries.
This guarantees that there exists at least one maxima inside the polygon.
\end{proof}

\begin{oldLemma}{l-gain-loss}
Suppose $\rgain$ is the fraction of susceptible population that became immune via vaccination, and $\rloss$ is the fraction of
immune population that lost immunity over a specified period of time. Then the new trajectory of the pandemic is obtained
by multiplying both $\beta$ and $\r$ (equivalently both $\bt$ and $\rt$) by $1+\frac{\rloss-\rgain}{\r}$.
\end{oldLemma}

\begin{proof}
During the course of the pandemic, first few phases did not have any vaccination or immunity loss. Consider the first phase
with either immunity loss or gain through vaccination or both. Suppose fraction of population that gains immunity through vaccination in this phase is $\rgain$ and the fraction of population that loses immunity is $\rloss$. Consider a time instant $t$ in the stable period of the phase when the changes in immunity have already taken place. Then the fraction of removed population would be $\R(t) + \rgain P_0 - \rloss P_0$ where $\R(t)$ is the fraction of removed population if there was no change in immunity levels. Therefore, the fundamental equation~(\ref{eq:NTdisc}) changes to:
\begin{eqnarray*}
\N_T(t+1) & = & \beta(1-\e) S\T \\
		& = & \beta(1-\e) (1 - \frac{1}{\r P_0}(\M+\R+\rgain P_0-\rloss P_0)) \T \\
		& = & \beta(1-\e) (1 - \frac{\rgain - \rloss}{\r} - \frac{1}{\r P_0} (\M+\R)) \T \\
		& = & \beta(1-\e) f (1 - \frac{1}{\r f P_0} (\M+\R)) \T \\
		& = & \beta(1-\e)f (1 - c - \frac{1}{\e\r f P_0} (\T+\R_T)) \T \\
		& = & \bt f (1 - \frac{1}{\rt f P_0} (\T + \R_T)) \T
\end{eqnarray*}
where $f = 1 - \frac{\rgain - \rloss}{\r}$. Therefore, fundamental equation now holds with values of $\beta$ and $\r$ multiplied by $f$. The other equations of the model (\ref{eq:Tdisc} and \ref{eq:disc}) can easily be seen to hold with the
same change in values of $\beta$ and $\r$.
\end{proof}

\begin{oldLemma}{l-new-phase}
Suppose a new phase begins at time $t_0$ with a drift period of $d$ days. Further, suppose that the value of parameter $\e$ changes from $\e_0$ to $\e_1$ during the new phase. Then,
$\M(t_0+d) = \frac{1}{\e_1} \T(t_0+d)$.
\end{oldLemma}

\begin{proof}
The model parameters $\beta(1-\e)$, $\e$, $\r$ and $1-c$ change multiplicatively until they stabilize to new values. Suppose that in one day, $\beta(1-\e)$ changes by a factor of $f_b$, $\e$ by a factor of $f_e$, $\r$ by a factor of $f_r$ and $1-c$ by a factor of $f_c$. Then, composite parameter $\bt = \beta(1-\e)(1-c)$ will change by a factor of $f_bf_c$ and 
$\rt = \e\r(1-c)$ will change by a factor of $f_ef_rf_c$. We know that after the drift period, $\bt$ and $\rt$ stabilize. Suppose that $\e$ continues to change even after the drift period of $d$ days and stabilizes after $D$ days in the phase. We now consider following cases:
\begin{itemize}
\item
$\bt$ changes during the drift period and $\beta(1-\e)$ changes on day $D-1$. In that case, change in $\bt$ on day $D-1$ equals $f_bf_c$ which is not equal to $1$ since $\bt$ changes during drift period. This is not possible.
\item
$\bt$ changes during the drift period and $\beta(1-\e)$ does not change on day $D-1$. Then change in $\bt$ on day $D-1$ equals $f_c$, and since $\bt$ does not change on day $D-1$, $f_c = 1$. Computation leading up to derivation of equation~(\ref{eq:x}), as in the proof of Theorem~\ref{t-finite-eps}, can be carried out for day $D$ (since all parameters stabilize by then), and substituting $y = \frac{1}{f_c} = 1$ in the equation~(\ref{eq:x}) we get $f_e = x = 1$ (note that $\rt_D = \rt_{D-1}$). This contradictions the assumption that $\e$ changes during the phase.
\item
$\rt$ changes during drift period and $\r$ changes on day $D-1$. Then change in $\rt$ on day $D-1$ equals $f_rf_ef_c$ which is not equal to $1$ since $\rt$ changes during drift period. This is not possible.
\item
$\rt$ changes during drift period and $\r$ does not change on day $D-1$. Then change in $\rt$ on day $D-1$ equals $f_ef_c$ which must be equal to $1$ since $\rt$ does not change after $d$ days. Therefore, $f_e = 1/f_c$. Going back to equation~(\ref{eq:x}) and substituting $x =f_e = 1/f_c = y$, we again get $f_e = 1 = f_c$, contradicting the assumption that $\e$ changes during the phase.
\end{itemize}
Together, the cases above cover all possibilities and hence we conclude that $D = d$ and therefore, 
$\M(t_0+d) = \frac{1}{\e_1} \T(t_0+d)$.
\end{proof}

\begin{oldLemma}{l-any-eps}
Given detected new cases trajectory, $\N_T(t)$, $0 \leq t \leq t_F$, there exist infinitely many total new cases trajectories and corresponding values of $\e$ consistent with $\N_T$.
\end{oldLemma}

\begin{proof}
Given $\N_T(t)$, $0\leq t\leq t_F$, we can compute phases of the trajectory and values of $\bt$ and $\rt$ for all phases as shown in section~\ref{sec-detected-parameters}, as well as $\C_T(t)$, $\T(t)$ and $\R_T(t)$ for the entire duration.

Choose any value $\e_0$ in the range $[0.9, 1.0]$. Fix $\e = \e_0$ and $c = 0$. This allows us to compute the values of $\beta$ and $\r$ for all phases (the value of $\r$ will be at most $\frac{1}{0.9}$ times the value of $\rt$ at any time).

Let $\N(t) = \frac{1}{\e} \N_T(t)$, and $\M(t) = \frac{1}{\e}\T(t)$ for $0 \leq t\leq t_F$. Setting $\R(0) = 0$, and using the equation~(\ref{eq:disc}) for $R$, we get that $\R(t) = \g \sum_{s = 0}^{t-1} \M(s) = \frac{\g}{\e} \sum_{s = 0}^{t-1} \T(s) = \frac{1}{\e} \R_T(t)$.
Then, for all $t$, $0\leq t\leq t_F$:
\begin{eqnarray*}
\N(t) & = & \frac{1}{\e} \N_T(t) \\
	& = & \frac{1}{\e} \bt (1 - \frac{1}{\rt P_0} (\T(t-1) + \R_T(t-1))) \T(t-1) \\
	& = & \bt (1 - \frac{1}{\r P_0} (\M(t-1) + \R(t-1))) \M(t-1) \\
	& = & \beta(1-\e) S(t-1) \M(t-1).
\end{eqnarray*}
It is straightforward to see that the remaining model equations (\ref{eq:disc}) are also satisfied. As there
are infinitely many values in the range $[0.9, 1.0]$, the proof is complete.
\end{proof}

\begin{oldTheorem}{t-finite-eps}
Given detected new cases trajectory, $\N_T(t)$, $0 \leq t \leq t_F$ and $\C(0)$, there exist only finitely many trajectories for $\N(t)$ consistent with $\N_T$. Further, a good estimate for all the trajectories can be obtained efficiently.
\end{oldTheorem}

\begin{proof}
Proof is by induction on the number of phases. In the base case we have only one phase. For this phase, there is no
drift period since there are no previous values of parameters. Therefore, parameter values stay the same throughout the phase duration. Let $\beta_1$, $\r_1$, $\e_1$ and $c_1$ be the parameter values governing the actual trajectory for this phase. Therefore, $\M(t) = \frac{1}{\e_1} \T(t)$ and $\R(t) = \frac{1}{\e_1} \R_T(t) + c_1\r_1 P_0$ for the entire phase.
Note that $\R(0) = 0$ since at time $t=0$, when the pandemic starts, there are no recoveries. Hence, $c_1 = 0$.
Further, $\e_1 = \T(0)/\M(0) = \T(0)/\C(0)$.
From this, we can compute
\begin{eqnarray*}
\r_1 & = & \frac{1}{\e_1} \rt_1 \\
\beta_1 & = & \bt_1 / (1-\e_1)
\end{eqnarray*}
giving values of all parameters for first phase, using which the trajectory can be computed for the first phase uniquely.

Suppose there are finitely trajectories up to phase $i-1$. Fix any one trajectory with values of four
parameters in phase $i-1$ being $\beta_{0}$, $\r_{0}$, $\e_{0}$ and $c_{0}$. Let $t_0$ be the time when
phase $i$ starts. We have:

\[ \begin{bmatrix} \M(t_0) \\ \C(t_0) \end{bmatrix} =  \frac{1}{\e_{0}} \begin{bmatrix} \T(t_0) \\ \C_T(t_0) \end{bmatrix} +
 c_0\r_{0} \begin{bmatrix} 0 \\ P_0  \end{bmatrix}.
\]

Suppose phase $i$ has a drift period of $d$ days. In the model, the parameter values change multiplicatively during the drift period. Let $\e_j = \e_{0} x^j$, and $1-c_{j} = (1-c_{0})/ y^j$ for $1\leq j\leq d$, where $x$ and $y$ are unknown multipliers by which the two parameters change every day.
The final value of the parameters will be $\e_d = \e_{0}x^{d}$ and $1-c_d = (1-c_{0})/ y^{d}$. 

Let $\bt_{j} = \bt_{0}(\frac{\bt_d}{\bt_{0}})^{j/d}$ and $\rt_{j} = \rt_{0}(\frac{\rt_d}{\rt_{0}})^{j/d}$,
for $1\leq j \leq d$. These numbers can be computed since $\bt_{0}$, $\bt_d$, $\rt_{0}$, and $\rt_d$ are known.

Let $\beta_{j} = \frac{\bt_{j}}{(1-\e_{j})(1-c_{j})}$ and $\r_{j} = \frac{\rt_{j}}{\e_{j}(1-c_{j})}$ for 
$1\leq j \leq d$. Then we can write:

\[ \begin{bmatrix} \M(t_0+j) \\ \R(t_0+j) \end{bmatrix}  =   \begin{bmatrix} g_j & 0 \\ \g & 1 \end{bmatrix} \cdot \begin{bmatrix} \M(t_0+j-1) \\ \R(t_0+j-1) \end{bmatrix} \]
where 
\begin{eqnarray*}
 g_j & = & \beta_{j-1}(1-\e_{j-1})(1-\frac{\M(t_0+j-1)+\R(t_0+j-1)}{\r_{j-1}P_0}) -\g+1  \\
	& = & \bt_{j-1}(\frac{1}{1-c_{j-1}} - \frac{\e_{j-1}}{\rt_{j-1}}\frac{\M(t_0+j-1)+\R(t_0+j-1)}{P_0}) -\g+1 \\
	& = & \bt_{j-1}(\frac{y^{j-1}}{1-c_{0}} - \frac{\e_{0}x^{j-1}}{\rt_{j-1}}\frac{\M(t_0+j-1)+\R(t_0+j-1)}{P_0})-\g+1
\end{eqnarray*}
Therefore, both $\M(t_0+j)$ and $\R(t_0+j)$ are polynomials in $x$ and $y$. It is straightforward to show that the degrees of $\M(t_0+j)$ and $\R(t_0+j)$ equal $2^j-j-1$ and $2^{j-1}-j-2$ respectively.

At the end of drift period, we have:
\begin{eqnarray}
\begin{bmatrix} \M(t_0+d) \\ \R(t_0+d) \end{bmatrix} & = & \frac{1}{\e_d} \begin{bmatrix} \T(t_0+d) \\ \R_T(t_0+d) \end{bmatrix} + c_d \r_d\begin{bmatrix}0 \\  P_0\end{bmatrix} \nonumber \\
				& = & \frac{1}{\e_d} \begin{bmatrix} \T(t_0+d) \\ \R_T(t_0+d) \end{bmatrix} + c_d\frac{\rt_d}{\e_d(1-c_d)} \begin{bmatrix} 0 \\ P_0\end{bmatrix}\nonumber \\
				& = & \frac{1}{\e_{0}x^{d}}\begin{bmatrix}\T(t_0+d) \\ \R_T(t_0+d) \end{bmatrix} 
				+ \frac{\rt_d}{\e_{0}x^{d}}(\frac{y^{d}}{1-c_{0}} - 1)\begin{bmatrix} 0 \\  P_0\end{bmatrix}\mbox{~~~~~~} \label{eq:81}
\end{eqnarray}

For what values of unknown multipliers $x$ and $y$ are the relationships in equation~(\ref{eq:81}) satisfied?
To see this, we analyze the quantities $\N(t_0+d)$, $\M(t_0+d)$, and $\C(t_0+d)$. From (\ref{eq:disc}) we have:
\begin{eqnarray*}
\C(t_0+d) & = & \C(t_0+d-1) + \N(t_0+d)  \\
\M(t_0+d) & = & \N(t_0+d) + (1-\g) \M(t_0+d-1) \\
\N(t_0+d) & = & \beta_{d-1}(1-\e_{d-1})S(t_0+d-1) \M(t_0+d-1) \\
		& = & \beta_{d-1}(1-\e_{d-1}) \left(1 - \frac{1}{\r_{d-1} P_0} \C(t_0+d-1)\right) \M(t_0+d-1)
\end{eqnarray*}
Therefore,
\begin{eqnarray*}
\C(t_0+d-1) & = & \frac{1}{\e_d} \C_T(t_0+d) + c_d\r_d P_0 - \frac{1}{\e_d} \N_T(t_0+d) \\
		& = & \frac{1}{\e_d} \C_T(t_0+d-1)  + c_d\r_d P_0 \\
\M(t_0+d-1) & = & \frac{1}{\e(1-\g)} (\T(t_0+d) - \N(t_0+d)) \\
			& = & \frac{1}{\e} \T(t_0+d-1) \\
\N(t_0+d) & = & \beta_{d-1}(1-\e_{d-1}) \left(1 - \frac{1}{\e_d\r_{d-1}P_0} \C_T(t_0+d-1) - \frac{c_d\r_d}{\r_{d-1}}\right) \frac{1}{\e_d} \T(t_0+d-1) \\
		& = & \frac{\bt_{d-1}}{\e_d(1-c_{d-1})} \left(1 - \frac{1}{\e_d\r_{d-1}P_0} \C_T(t_0+d-1) - \frac{c_d\r_d}{\r_{d-1}}\right) \T(t_0+d-1).
\end{eqnarray*}
Since
\[ \N(t_0+d) = \frac{1}{\e_d} \N_T(t_0+d) = \frac{\bt_{d-1}}{\e_d}\left(1 - \frac{1}{\rt_{d-1} P_0} \C_T(t_0+d-1)\right) \T(t_0+d-1) \]
where the second equality is from fundamental equation (\ref{eq:NTdisc}), we have:
\begin{eqnarray*}
1 - \frac{1}{\rt_{d-1} P_0} \C_T(t_0+d-1) & = & \frac{1}{1-c_{d-1}} \left(1 - \frac{1}{\e_d\r_{d-1}P_0} \C_T(t_0+d-1) - \frac{c_d\r_d}{\r_{d-1}}\right) \\
	& = & \frac{1}{1 - c_{d-1}} - \frac{c_d\r_d}{(1-c_{d-1})\r_{d-1}} - \frac{1}{\e_d\r_{d-1}(1-c_{d-1})P_0} \C_T(t_0+d-1) \\
	& = & \frac{y^{d-1}}{1-c_0} - \frac{\e_{d-1}c_d\r_d}{\rt_{d-1}} - \frac{1}{x\rt_{d-1}P_0}\C_T(t_0+d-1) \\
	& = & \frac{y^{d-1}}{1-c_0} - \frac{c_d\rt_d}{x(1-c_d)\rt_{d-1}} - \frac{1}{x\rt_{d-1}P_0}\C_T(t_0+d-1) \\
	& = & \frac{y^{d-1}}{1-c_0} + \frac{\rt_d}{x\rt_{d-1}} - \frac{y^d\rt_d}{x(1-c_0)\rt_{d-1}}  - \frac{1}{x\rt_{d-1}P_0}\C_T(t_0+d-1).
\end{eqnarray*}
Therefore,
\begin{eqnarray}
x & = & \frac{\left(\frac{\rt_d}{\rt_{d-1}} - \frac{y^d\rt_d}{(1-c_0)\rt_{d-1}}  - \frac{1}{\rt_{d-1}P_0}\C_T(t_0+d-1)\right)}{\left(1 - \frac{y^{d-1}}{1-c_0} - \frac{1}{\rt_{d-1} P_0}\C_T(t_0+d-1)\right)}. \label{eq:x}
\end{eqnarray}
Equation~(\ref{eq:x}) expresses $x$ as a degree $d$ rational function of $y$. Substituting this in the polynomial 
$\M(t_0+d)$, we obtain a degree $d(2^d-d-1)$ rational function in $y$. Equating it to $\frac{1}{\e_d}\T(t_0+d) = \frac{1}{\e_0 x^d}\T(t_0+d)$ results in a polynomial in $y$ of degree bounded by $d(2^d-d)$, say $P(y)$, whose roots are the possible values of $y$. Therefore, there are at most $d(2^d-d)$ many values of $(x,y)$ that satisfy all the required equations. 

While the potential number of solutions is very large for even moderate values of $d$ (say $d > 20$), most of the solutions are likely to be infeasible since a feasible solution needs to satisfy the conditions that $\e_d$ must be in the range $[0,1]$ and $c_d$ in the range $[-1, 1]$.
Further, there is a simple and efficient way to list out good estimates for all the solutions: since $-1 < c_d < 1$, one can step through possible values of $c_d$ in the range using small discrete steps, compute the value of $y$ for the chosen value of $c_d$,  compute value of $x$ using equation~(\ref{eq:x}), and then check if $P(y)$ is close to zero. This will list out good estimates of all feasible solutions.
\end{proof}

\end{document}